\documentclass[twocolumn,superscriptaddress,amsmath,amssymb,aps,pra,floatfix]{revtex4-1}

\usepackage[utf8]{inputenc}
\usepackage{wrapfig}
\usepackage{float}
\usepackage{amssymb}
\usepackage{amsfonts}
\usepackage{amsmath}
\usepackage{cases}
\usepackage{stmaryrd}\usepackage{tipa}
\usepackage[dvips]{graphicx}
\usepackage{dcolumn}
\usepackage{bbold}
\usepackage{graphicx}
\usepackage{graphicx}
\usepackage{subfigure}
\usepackage{dcolumn}
\usepackage{bm}
\usepackage{color}
\usepackage{xcolor}
\usepackage{CJK}
\usepackage{subfigure}
\usepackage{physics}
\usepackage{array}
\usepackage{multirow}
\usepackage{nopageno}
\usepackage{mathtools}
\usepackage{enumerate}
\usepackage{comment}
\usepackage[colorlinks=true, linkcolor=red, citecolor=blue]{hyperref}

\definecolor{gg}{RGB}{8, 135, 68}
\newcommand{\mathbbm}[1]{\text{\usefont{U}{bbm}{m}{n}#1}}
\newcommand{\Id}{\mathbbm{1}} 
\DeclareMathOperator{\diff}{d}

\begin{document}

\title{Open-Air Microwave Entanglement Distribution for Quantum Teleportation}

\author{Tasio Gonzalez-Raya}
\email{tasio.gonzalez@ehu.eus}
\affiliation{Department of Physical Chemistry, University of the Basque Country UPV/EHU, Apartado 644, 48080 Bilbao, Spain}
\affiliation{EHU Quantum Center, University of the Basque Country UPV/EHU, Bilbao, Spain}
\author{Mateo Casariego}
\affiliation{Instituto de Telecomunicações, Physics of Information and Quantum Technologies Group, Portugal}
\affiliation{Instituto Superior Técnico, Universidade de Lisboa, Portugal}
\author{Florian Fesquet}
\affiliation{Walther-Mei\ss{}ner-Institut, Bayerische Akademie der Wissenschaften, 85748 Garching, Germany}
\affiliation{Physik-Department, Technische Universit{\"a}t M{\"u}nchen, 85748 Garching, Germany}
\author{Michael Renger}
\affiliation{Walther-Mei\ss{}ner-Institut, Bayerische Akademie der Wissenschaften, 85748 Garching, Germany}
\affiliation{Physik-Department, Technische Universit{\"a}t M{\"u}nchen, 85748 Garching, Germany}
\author{Vahid Salari}
\affiliation{Department of Physical Chemistry, University of the Basque Country UPV/EHU, Apartado 644, 48080 Bilbao, Spain}
\affiliation{Basque Center for Applied Mathematics (BCAM), Alameda de Mazarredo 14, 48009 Bilbao, Basque Country, Spain}
\author{Mikko M\"{o}tt\"{o}nen}
\affiliation{IQM, Keilaranta 19, 02150 Espoo, Finland}
\affiliation{QCD Labs, QTF Centre of Excellence, Department of Applied Physics, Aalto University, P.O. Box 13500, FI-00076 Aalto, Finland}
\affiliation{VTT Technical Research Centre of Finland Ltd. \& QTF Centre of Excellence, P.O. Box 1000, 02044 VTT, Fin- land}
\author{Yasser Omar}
\affiliation{Instituto de Telecomunicações, Physics of Information and Quantum Technologies Group,  Portugal}
\affiliation{Instituto Superior Técnico, Universidade de Lisboa, Portugal}
\affiliation{Portuguese Quantum Institute, Portugal}
\author{Frank Deppe}
\affiliation{Walther-Mei\ss{}ner-Institut, Bayerische Akademie der Wissenschaften, 85748 Garching, Germany}
\affiliation{Physik-Department, Technische Universit{\"a}t M{\"u}nchen, 85748 Garching, Germany}
\affiliation{Munich Center for Quantum Science and Technology (MCQST), Schellingstr. 4, 80799 Munich, Germany}
\author{Kirill G. Fedorov}
\affiliation{Walther-Mei\ss{}ner-Institut, Bayerische Akademie der Wissenschaften, 85748 Garching, Germany}
\affiliation{Physik-Department, Technische Universit{\"a}t M{\"u}nchen, 85748 Garching, Germany}
\author{Mikel Sanz}
\email{mikel.sanz@ehu.eus}
\affiliation{Department of Physical Chemistry, University of the Basque Country UPV/EHU, Apartado 644, 48080 Bilbao, Spain}
\affiliation{EHU Quantum Center, University of the Basque Country UPV/EHU, Bilbao, Spain}
\affiliation{Basque Center for Applied Mathematics (BCAM), Alameda de Mazarredo 14, 48009 Bilbao, Basque Country, Spain}
\affiliation{IKERBASQUE, Basque Foundation for Science, Plaza Euskadi 5, 48009 Bilbao, Spain}

\begin{abstract}
Microwave technology plays a central role in current wireless communications, standing among them mobile communication and local area networks (LANs). The microwave range shows relevant advantages with respect to other frequencies in open-air transmission, such as low absorption losses and low energy consumption, and it is additionally the natural working frequency in superconducting quantum technologies. Entanglement distribution between separate parties is at the core of secure quantum communications. Therefore, understanding its limitations in realistic open-air settings, specially in the rather unexplored microwave regime, is crucial for transforming microwave quantum communications into a mainstream technology. Here, we investigate the feasibility of an open-air entanglement distribution scheme with microwave two-mode squeezed states. First, we study the reach of direct entanglement transmission in open-air, obtaining a maximum distance of approximately 500 meters in a realistic setting with state-of-the-art experimental parameters. Afterwards, we adapt entanglement distillation and entanglement swapping protocols to microwave technology in order to reduce environmental entanglement degradation. While entanglement distillation helps to increase quantum correlations in the short-distance low-squeezing regime by up to $46\%$, entanglement swapping increases the reach by $14\%$. Then, we compute the fidelity of a continuous-variable quantum teleportation protocol using open-air-distributed entanglement as a resource. Finally, we adapt the machinery to explore the limitations of quantum communication between satellites, where the thermal noise impact is substantially reduced and diffraction losses are dominant.
\end{abstract}

\maketitle

\section{Introduction}\label{section:introduction}

Quantum communication~\cite{Gisin2007,Yuan2010,Krenn2016} represents an application of quantum information theory that tackles the subject of information transfer by taking advantage of purely quantum resources, such as superposition and entanglement. By establishing quantum channels to share quantum resources, namely quantum states, as well as secure classical channels, it aims at outperforming classical communication protocols in both efficiency and security. 

Among the best-known quantum communication protocols, quantum teleportation~\cite{Bennett1993,Pirandola2015} aims at transferring information of an unknown quantum state held by one party, to a second party at a remote location, by means of an entangled resource and classical communication. Initially proposed for discrete-variable quantum states~\cite{Baur2012,Steffen2013}, this protocol has also been studied in continuous-variable settings~\cite{Pirandola2006,Braunstein1998,Fedorov2021}.

Another notorious advantage of quantum communication is quantum key distribution~\cite{Diamanti2016,Wang2021}, whose foundation has been set by two distinct protocols: BB84~\cite{Bennett1984} and E91~\cite{Eckert1991}. These protocols allow two distant parties to develop a shared random key, which is unconditionally secure against eavesdroppers by virtue of quantum laws. 

At the heart of many quantum communication protocols lies entanglement distribution~\cite{Cirac1997,Chou2007,Mista2009,Dias2020}, also a key point for the famous quantum internet idea~\cite{Kimble2008,Wehner2018,Gyongyosi2019}; quantum entanglement distribution represents the act of sharing entangled states between communication parties. This has been experimentally attained~\cite{Herbst2015,Wengerowsky2019,Yin2017}, as well as quantum key distribution~\cite{Liao2017,Liao2018,Minder2019} and quantum teleportation~\cite{Bouwmeester1997,Furusawa1998,Boschi1998,Jin2010,Ma2012,Ren2017}, through optical fibers and through open-air. 

The subject of quantum communication has been developped parallel to other areas such as quantum computing, quantum sensing~\cite{Degen2017,Pirandola2018} or quantum metrology~\cite{Giovannetti2011,Polino2020}, to which it is usually tangential. Quantum computing, for example, is expected to benefit from efficient transfer of quantum information between processing units, a proposal that is categorized as distributed quantum computing~\cite{Rodrigo2020,DiAdamo2021}. Quantum sensing and quantum metrology profit from the use of entanglement in their attempt to perform high-resolution measurements on various systems. By using quantum resources, they can reach ultimate measurement limits set by quantum mechanics, thus outperfoming classical strategies. Respective examples can be found in detection of gravitational waves at LIGO~\cite{LIGO2011,Oelker2016}, in measurement of biological systems~\cite{Taylor2016,Mauranyapin2017}, quantum imaging~\cite{Moreau2019,Berchera2019}, navigation~\cite{Fink2019}, and synchronization~\cite{Quan2016}, among others.

The flourishing of these quantum-information-based fields has come hand in hand with the development of quantum technologies, among which superconducting circuits stand out in terms of controllability, scalability, and coherence. Partially, this is due to the development of Josephson junctions (JJs)~\cite{Josephson1969}, nonlinear elements with essential applications in quantum computation~\cite{Buttiker1987,Bouchiat1998,Koch2007} and quantum information processing~\cite{Yamamoto2007,Castellano-Beltran2008}. This has led to different experiments in quantum state transfer and remote entanglement preparation between various JJ-based superconducting devices~\cite{Axline2018, Campagne2018, Kurpiers2018, Leung2019, Roch2015, Narla2016,Dickel2018,Magnard2020}, as well as to sensitive noise analysis~\cite{Goetz2017, Goetz2017_2}. These devices naturally work at microwave frequencies ($1$-$100$ GHz), for which the number of thermal photons per mode at room temperature ($T=300$ K) is around $1250$ at $5$ GHz, thus creating the need for cryogenic cooling in order to shield superconducting circuits from thermal noise. This problem is somewhat non-existing in the optical regime, where most quantum communication experiments have been performed so far. However, in this regime there are many other sources of errors and inefficiencies~\cite{Sanz2018}: large absorption losses in open-air and significant power consumption requirements. At the same time, in order to establish a quantum communication channel between superconducting quantum circuits, one requires either to convert microwave photons to optical domain~\cite{Forsch2020, Rueda2019} or to use microwave quantum signals directly. The former approach still suffers from huge conversion quantum inefficiencies on the order of $10^{-5}$. Therefore, it is natural to consider the purely microwave quantum communication approach, its advantages and limitations. 

Another interesting application of the Josephson junction is the Josephson parametric amplifier (JPA), a device that can generate squeezed states~\cite{Fedorov2016}, from which entangled resources~\cite{Pogorzalek2019,Fedorov2018} can be produced for quantum communication with microwaves. Among the various applications of open-air entanglement distribution, quantum teleportation represents one of the fundamental quantum communication protocols, since its associated network has just two nodes: Alice and Bob. This protocol has been explored in the microwave cryogenic environment, both theoretically~\cite{DiCandia2015} and experimentally~\cite{Fedorov2021}. However, a realistic model for open-air quantum teleportation with microwaves should take into account additional challenges associated with impedance mismatches and absorption losses in order to find ways to mitigate these imperfections.

In this article, we address two pragmatic questions: which is the maximum distance for open-air microwave Gaussian entanglement distribution in a realistic scenario, and which technological and engineering challenges remain to be faced. In particular, we adapt the Braunstein-Kimble quantum teleportation protocol employing entangled resources previously distributed through open-air, adapted to microwave technology. The possibility of performing this protocol is caused by the recent breakthrough in the development of microwave homodyning~\cite{Fedorov2021} and photocounting~\cite{Dassonneville2020} schemes. When formulated in continuous variables (CVs), teleportation assumes a previously shared entangled state, ideally a two-mode squeezed vacuum (TMSV) state with infinite squeezing. In real-life, however, only a finite squeezing level can be produced, making the state sensitive to entanglement degradation whenever either one or both modes, are exposed to decoherence processes like thermal noise and/or photon losses. As the thermal microwave background in space is smaller, we investigate the distances for entanglement preservation in microwave quantum communication between satellites. Based on previous works~\cite{Pirandola2021,Pirandola2021_2}, we neglect environmental attenuation and focus on diffraction losses, a powerful loss mechanism in microwaves. The article is organized as follows: In section~\ref{section:quantum_CV}, we introduce the main concepts in quantum CVs, which are needed for describing quantum microwaves, and briefly discuss the quantum teleportation protocol of Braunstein and Kimble~\cite{Braunstein1998}. In section~\ref{section:open_air_entanglement_distribution}, we study the generation of two-mode squeezed states and the challenges of their subsequent ditribution through open-air, and compute maximum distances of entanglement preservation for various physical situations. In section~\ref{subsection:distillation_and_swapping}, we review entanglement distillation and entanglement swapping techniques for reducing the effects of noise and losses in open-air. In section~\ref{section:homodyning_and_photocounting} we consider recent advances in microwave photodetection and homodyning, and address their current limitations. In section~\ref{section:teleportation}, we investigate open-air microwave quantum teleportation fidelities using the various quantum states derived in the manuscript, and we conclude by addressing the same concerns in quantum communication between satellites in section~\ref{section:satellites}. 

\section{Quantum Continuous Variable Formalism}\label{section:quantum_CV}
\subsection{Review of Gaussian states}
When some degree of freedom of a quantum system is described by a continuous-spectrum operator, we say that it is a `continuous variable' (CV). Bosonic CV states are the ones whose quadratures (or, equivalently, their creation and annihilation operators) have a continuous spectrum or, equivalently, the complete description of the Hilbert space requires an infinite-dimensional basis (typically the Fock basis). Gaussian states are CV states associated with Hamiltonians that are, at most, quadratic in the field operators. As such, their full description does not require the infinite-dimensional density matrix, and can be compressed into a vector and a matrix, called the displacement vector and the covariance matrix, respectively. These are related to the first and second moments of a Gaussian distribution, hence their name ``Gaussian states". For a system with density matrix $\rho$ describing $N$ distinguishable modes, or particles, the displacement vector $\bm{d}$ is a $2N$-vector and the covariance matrix $\Sigma$ is a $2N\times 2N$ square matrix: 
\begin{align}
\bm{d} &:=\Tr\left[ \rho \hat{\bm{r}}\right]\\
\Sigma &:= \Tr\left[ \rho \lbrace (\hat{\bm{r}} - \bm{d}), (\hat{\bm{r}} - \bm{d})^\intercal\rbrace\right],
\end{align}
where $\hat{\bm{r}}:= ( \hat{x}_1, \hat{p}_1, \hat{x}_2, \hat{p}_2, \ldots, \hat{x}_N, \hat{p}_N)$ defines the so-called `real basis', for which canonical commutation relations read: $\left[\hat{\bm{r}}, \hat{\bm{r}}^\intercal\right] = i\Omega$, where $\Omega = \bigoplus_{j=1}^N \Omega_1$ is the quadratic (or symplectic) form, and 
\begin{equation}
 \Omega_1 =
    \begin{pmatrix}
       0 & 1\\
       -1 & 0
    \end{pmatrix},
   \end{equation}
where we have chosen natural units, $\hbar = 1$. Note that the \textit{canonical} position and momentum operators  are defined by the choice $\kappa=2^{-1/2}$ in $\hat{a}_j = \kappa(\hat{x}_j + i \hat{p}_j)$.

The normal mode decomposition theorem~\cite{Arvind1995}, which follows from Williamson's seminal work \cite{Williamson1939}, can be stated as: every positive-definite Hermitian matrix $\Sigma$ of dimension $2N\times2N$ can be diagonalized with a symplectic matrix $S$: $D = S \Sigma S^\intercal$, with $D = \text{diag}\left( \nu_1, \nu_1,  \ldots, \nu_N, \nu_N \right)$, where $\nu_{a}$ are the symplectic eigenvalues of $\Sigma$, defined as the positive eigenvalues of matrix $i\Omega\Sigma$. A Gaussian state satisfies $\nu_{a} \geq 1$, with equality for all $a$ strictly for the pure state case (which meets $\det\Sigma=1$). As a measure of bipartite, mixed state entanglement, the negativity is the most commonly used entanglement monotone, and is defined as $2\mathcal{N}(\rho):=\norm{\tilde{\rho}}_1-1$, where $\norm{\tilde{\rho}}_{1}:=\Tr\sqrt{\tilde{\rho}^\dagger \tilde{\rho}}$ is the trace norm of the partially transposed density operator. In general, $\mathcal{N}(\rho) =\abs{ \sum_j \lambda_j}$ with $\lambda_j$ the negative eigenvalues of $\tilde{\rho}$. For a bipartite Gaussian state with covariance matrix 
\begin{equation}\label{eq:twoMode}
\Sigma = \begin{pmatrix}
\Sigma_A & \varepsilon_{AB}\\
\varepsilon_{AB}^\intercal & \Sigma_B
\end{pmatrix},
\end{equation}
one defines the two partially-transposed symplectic eigenvalues as
\begin{equation}\label{eq:ptseigen}
    \tilde{\nu}_{\mp}:=\sqrt{\frac{\tilde{\Delta} \mp \sqrt{\tilde{\Delta}^2- 4 \det\Sigma}}{2}},
\end{equation}
where the partially-transposed symplectic invariant is $\tilde{\Delta} = \det \Sigma_A + \det \Sigma_B - 2\det \varepsilon_{AB}$. The negativity can then be obtained as
\begin{equation}\label{eq:negativity}
\mathcal{N}(\rho) =\max \left\{0, \frac{1-\tilde{\nu}_{-}}{2\tilde{\nu}_{-}}\right\}.
\end{equation} 
Hence, a bipartite Gaussian state is separable when the smaller partially-transposed symplectic eigenvalue meets the condition $\tilde{\nu}_{-} \geq 1$. Alternatively, it is entangled when $\tilde{\nu}_{-} < 1$ is met.

Coherent states $\lbrace\ket{\alpha}\rbrace_{\alpha\in \mathbb{C}}$ are defined as the eigenstates of the annihilation operator $\hat{a}$ with eigenvalue $\sqrt{2}\alpha =x + ip$ where $x, p\in \mathbb{R}$ are the eigenvalues of the canonical position and momentum  operators, respectively. They play an important role in quantum CVs, as they allow for a straighforward phase space description of Gaussian states. The displacement operator $\hat{D}_{-\bm{d}}:=\exp[-i\bm{d}\Omega_1\hat{\bm{d}}^\intercal]=\hat{D}_\alpha = \exp \left[ \alpha \hat{a}^\dagger - \bar{\alpha} \hat{a}\right]$ acts on the vacuum as $\hat{D}_\alpha\ket{0} = \ket{\alpha}$, and satisfies $\hat{D}_\alpha^\dagger = \hat{D}_{-\alpha}$. Coherent states are not orthogonal, and their overlap can be computed as $\bra{\beta}\ket{\alpha} = \exp\left[ -1/2(\alpha\bar{\beta} -\bar{\alpha}\beta -\abs{\alpha-\beta}^2\right]$. This does not prevent the set of all coherent states from forming a basis, which, though overcomplete, allows one to find the coherent states resolution of the identity: $\Id = \pi^{-1}\int \diff^2 \alpha \ket{\alpha}\bra{\alpha}$, where $\diff^2 \alpha \equiv \diff \Re{\alpha} \diff \Im{\alpha}$, enabling the computation of  traces of operators in an integral fashion: $\Tr [ \hat{O}] =\pi^{-1} \int \diff^2 \alpha \bra{\alpha}\hat{O}\ket{\alpha}$. In this context, it is common to use the fact that for a coherent state $\bm{d} = (x, p) = \sqrt{2}(\Re{\alpha}, \Im{\alpha})$, and so $2 \diff^2 \alpha  = \diff x \diff p$. More generally, an $n$-mode displacement operator may be defined via $\hat{D}_{-\bm{\mathrm{d}}} =\bigotimes_{j=1}^n  \hat{D}_{-\bm{d}_j} =\hat{D}_{-{\bigoplus_{j=0}^n}\bm{d}_j} $, where $\bm{d}_j:=(x_j, p_j)$, and $\Omega :=\bigoplus_{j=1}^n \Omega_1$.
A complete representation of states that is closely related to coherent states is given by the (Wigner) characteristic function, normally referred to simply as the characteristic function (CF), and that for an $n$-mode state $\rho$ (not necessarily Gaussian) is given by 
\begin{equation}
\chi(\bm{\mathrm{d}}) = \Tr\left[ \rho\hat{D}_{-\bm{\mathrm{d}}}\right],
\end{equation}
with normalization condition given by $\chi(\bm{0}) =1$. A Gaussian state of first and second moments $(\bm{\mathrm{d}}, \Sigma)$ has a CF given by
\begin{equation}
\chi_G(\bm{\mathrm{r}}) = e^{-\frac{1}{4}\bm{\mathrm{r}} \Omega \Sigma \Omega^\intercal \bm{\mathrm{r}}^\intercal}e^{-i\bm{\mathrm{r}} \Omega \bm{\mathrm{d}}^\intercal},
\end{equation}
where $\bm{\mathrm{r}} = (x_1, p_1, \ldots, x_N, p_N) \in \mathbb{R}^{2N}$.

\subsection{Quantum teleportation with CVs}
Quantum teleportation is a quantum communication protocol which, in principle, allows to achieve perfect transfer of quantum information between two parties by means of previously shared entanglement, combined with local operations and classical communication. The protocol was first proposed in 1993 by Bennett and collaborators~\cite{Bennett1993}, as a way to take advantage of an entangled resource for the task of sending an unknown quantum state from one place to another, using discrete-variable quantum states. The original idea was simple, yet powerful: assuming that a maximally entangled, bipartite Bell state was shared between two parties (Alice and Bob) prior to the start of the protocol, Alice, in possession of some \textit{unknown} state $\ket{\psi}=\alpha\ket{0} + e^{i\beta}\sqrt{1-|\alpha|^2}\ket{1}$ couples her part of the Bell state to $\ket{\psi}$ by means of a Bell measurement, whose 2-bit output she communicates classically to Bob. Upon receiving the message, Bob performs a conditional unitary on his part of the shared Bell state, recovering $\ket{\psi}$ modulo a global phase in his location.

A year later, Vaidman extended the idea to the transmission of a CV state by means of a perfectly correlated (singular) position-momentum EPR state shared by Alice and Bob~\cite{Vaidman1994}. In 1998, Braunstein and Kimble~\cite{Braunstein1998} made this idea more realistic by relaxing the correlation condition to more experimentally-accessible states, such as finitely-squeezed states. Their protocol, known as the  Braunstein-Kimble protocol, was first realised by A. Furusawa and collaborators in 1998 in the optical domain~\cite{Furusawa1998}. We shall review the protocol here for convenience.

Kimble and Braunstein derived an expression for fidelity between an unknown state of a single mode Bosonic field and a teleported copy, when imperfect quantum entanglement is shared between the two parties. A generalization to a broadband version, where the modes have finite bandwidths, followed quite directly~\cite{Braunstein2005}. In the Braunstein-Kimble protocol, Alice and Bob share a TMSV state, which enables them to teleport the complete state of a single mode of the electromagnetic field, where two orthogonal field quadratures play the role of position and momentum. Shortly after, quantum teleportation of an unknown coherent state was demonstrated, showing an average fidelity (see Eq. \eqref{eq:Pure-PS-fidelity})  $\overline{F}=0.58 \pm 0.02$~\cite{Furusawa1998}, which beat the maximum classical fidelity of $\overline{F} = 0.5$ for Gaussian states~\cite{Braunstein2005, Serafini2017, Weedbrook2012}. Other works followed, where the Bell measurement of two orthogonal quadratures was replaced by photon-number difference and phase sum, and the question of an optimal quantum teleportation protocol depending on the entangled resource was raised~\cite{Milburn1999}. Subtraction of single photons from TMS states has been shown to enhance the fidelity of teleportation \cite{Opatrny2000, Cochrane2002}. We shall review here the Braunstein-Kimble protocol, replacing the Wigner function approach with its Fourier transform, the characteristic function. 
\begin{figure}[t]
\includegraphics[width=0.3 \textwidth]{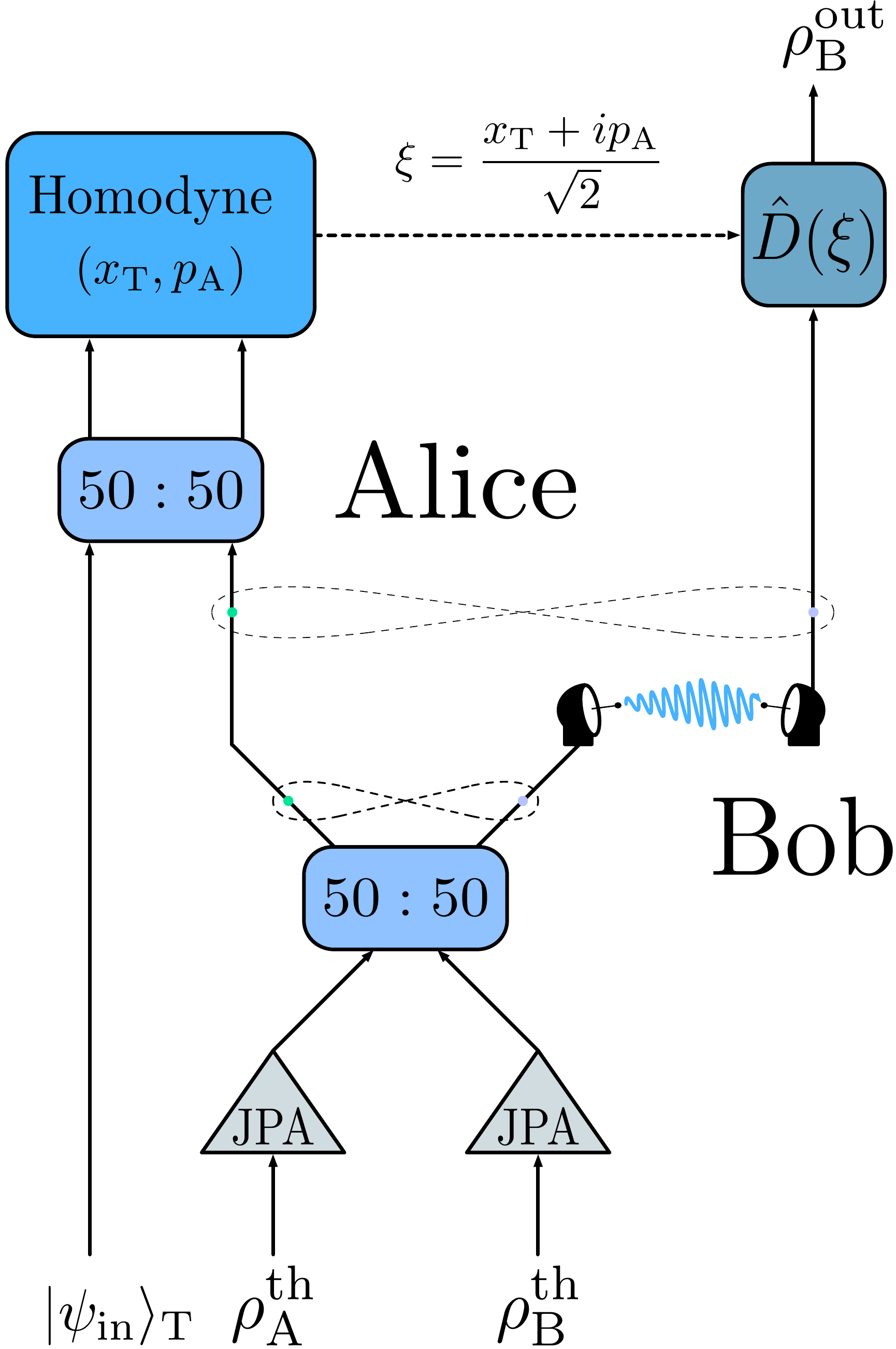}
\caption{Circuit representation of a CV microwave quantum teleportation protocol with Gaussian states. The entangled resource is harvested from two single-mode squeezed thermal states, generated from identical JPAs, which are then combined on a balanced beamsplitter. Assuming this state is generated by Alice, one of the modes has to be sent to Bob, represented here by the presence of antennae, in order for the two parties to share the entangled resource. Following this, Alice combined the target state to be teleported $|\psi_{\text{in}}\rangle_{\text{T}}$ with the mode of the entangled state she holds in a balanced beamsplitter, which is then subject to two homodyne detections, $x_\text{T}$ and $p_\text{A}$. The measurement results $\xi$ are communicated to Bob, who applies a displacement $\hat{D}(\xi)$ on his part of the entangled resource, resulting in the state $\rho^{\text{out}}_{\text{B}}$.}
\label{fig1}
\end{figure}
The protocol goes as follows: 
\begin{enumerate}
\item Alice uses a 50:50 beam splitter to couple her part of the resource state $\rho_{AB}$ with an incoming unknown state $\rho^{\text{in}}_{T}$. The output Hilbert spaces of this beam splitter are labeled $A$ and $T$.
\item Alice performs two homodyne detections, where each of the local oscillator phases are set in order to measure photocurrents, whose differences are integrated over some time, and proportional to quadratures $\hat{x}_T := (\hat{x}_1  + \hat{x}_\text{in})/\sqrt{2}$ and $\hat{p}_A := (\hat{p}_1  - \hat{p}_\text{in})/\sqrt{2}$. She sends the outcomes $(x_T, p_A)$ to Bob via a classical communication channel.
\item Bob, upon reception of the signal $(x_T, p_A)$, performs a displacement $\hat{D}(\xi)$ to his part of $\rho_{AB}$, with $\xi := (x_T + ip_A)/\sqrt{2}$. The state at Bob's location is now, in average, closer to $\rho_{\text{in}}$ than what it would be if no entanglement was present in $\rho_{AB}$.
\end{enumerate}
For simplicity, we will denote $(x_T, p_A) = (x,p)$. The conditional state that Bob has after knowing the outcomes of Alice's homodyne measurements is
\begin{equation}
\rho_{B}(x,p) = \frac{1}{P_{B}(x,p)} \langle \Pi(x,p)| \rho_{T}^{\text{in}}\otimes \rho_{AB}|\Pi(x,p)\rangle_{TA},
\end{equation}
with $P_{B}(x,p) = \Tr_{B}\langle \Pi(x,p)| \rho_{\text{in}}\otimes \rho_{AB}|\Pi(x,p)\rangle_{TA}$, and 
\begin{equation}
\ket{\Pi(x,p)}_{TA} = \frac{1}{\sqrt{2\pi}}\int_\mathbb{R}\diff y e^{ipy}\ket{x+y}_T\ket{y}_A,
\end{equation}
which is an element of the maximally entangled basis corresponding to Alice's Bell-like measurement. Now, this expectation value over the teleported ($T$) and the senders ($A$) modes is computed as 
\begin{widetext}
\begin{equation}
\langle \Pi(x,p)| \rho_{\text{in}}\otimes \rho_{AB}|\Pi(x,p)\rangle_{TA} = \frac{1}{2\pi}\int_{-\infty}^{\infty}\int_{-\infty}^{\infty} \diff y\diff y' e^{ip(y-y')}\langle x+y'|\rho_{\text{in}}|x+y\rangle_{T} \langle y'|\rho_{AB}| y\rangle_{A}.
\end{equation}
\end{widetext}
Once we have computed $\rho_{B}(x,p)$, we need to compute the outcoming state after the receiver applies the displacements, and average over all possible measurment outcomes
\begin{equation}
\rho_{B}^{\text{out}} = \int_{-\infty}^{\infty} \diff x \int_{-\infty}^{\infty} \diff p P_{B}(x,p)\hat{D}_{B}\left(\xi\right)\rho_{B}(x,p)\hat{D}^{\dagger}_{B}\left(\xi\right).
\end{equation}
As a measure of the quality of the protocol, one typically uses an overlap fidelity $F(\rho_\text{in}, \rho_\text{out}) = \Tr[\rho_{\text{in}}\rho_{\text{out}}]$, which corresponds to the Uhlmann fidelity $\left(\Tr[\sqrt{\sqrt{\rho_{\text{in}}}\rho_{\text{out}}\sqrt{\rho_{\text{in}}}}]\right)^{2}$ in the case when $\rho_{\text{in}}$ is pure. 

The figure of merit in quantum teleportation is the \textit{average} fidelity, which refers to the fact that we have averaged over all possible measurement outcomes,
\begin{equation}\label{eq:Pure-PS-fidelity}
\overline{F} = \Tr[\rho_{T}^{\text{in}}\rho_{B}^{\text{out}}].
\end{equation}
Sometimes it can be useful to have it written in terms of the CFs~\cite{Marian2006}:
\begin{equation}
\overline{F} =  \frac{1}{\pi}\int \diff ^{2}\beta \chi_{T}^{\text{in}}(-\beta)\chi_{B}^{\text{out}}(\beta),
\end{equation}
where the average over $(x,p)$ has already been performed in $\rho_{B}^{\text{out}}$.

If the resource state $\rho_{AB}$ is a Gaussian state with the covariance matrix given in Eq.~\eqref{eq:twoMode}, and the teleported state is a coherent state $|\alpha_{0}\rangle\langle\alpha_{0}|$, the average fidelity can be written as
\begin{equation}\label{eq:Gaussian-fidelity}
\overline{F} = \frac{1}{\sqrt{\det[\mathbb{1}_{2} + \frac{1}{2}\Gamma]}},
\end{equation}
with $\Gamma \equiv (\sigma_Z \Sigma_A\sigma_Z + \Sigma_B -\sigma_Z \varepsilon_{AB}-\varepsilon_{AB}^\intercal\sigma_Z)$. Coherent states are typically the ones chosen to be teleported due to the ease of their experimental generation. In theory, the result of the average fidelity does not depend on the displacement $\alpha_{0}$; therefore, it will suffice to use an unknown coherent state for a demonstration of quantum teleportation. In experiments, however, the teleportation fidelity may depend on $\alpha_{0}$. 

It is also interesting to see the average fidelity of a process in which $k$ teleportation protocols are concatenated:
\begin{equation}
\overline{F}^{(k)} = \frac{1}{\sqrt{\det[\mathbb{1}_{2} + \left(k-\frac{1}{2}\right)\Gamma)]}},
\end{equation}
assuming that, in each step, an entangled Gaussian resource with the covariance matrix that characterizes $\Gamma$ is used. 

Consider a symmetric covariance matrix with $\Sigma_{A}=\Sigma_{B}=\alpha\Id_{2}$ and $\varepsilon_{AB}=\gamma\sigma_{Z}$. Then, we have $\Gamma = 2(\alpha-\gamma)\Id_{2}$ and $\tilde{\nu}_{-}=\alpha-\gamma$, which leads to
\begin{equation}\label{eq:Gaussian-fidelity}
\overline{F} = \frac{1}{1+\tilde{\nu}_{-}}.
\end{equation}
It is easy to check the two following limits for the average teleportation fidelity of an arbitrary coherent state: $\lim_{\tilde{\nu}_{-}\rightarrow 1} \overline{F} = 1/2$ and $\lim_{\tilde{\nu}_{-}\rightarrow 0} \overline{F} = 1$. The first limit corresponds to using no entanglement ($\tilde{\nu}_{-}\geq1$), and is interpreted as the `classical teleportation' threshold, meaning that any approach giving an average fidelity of 0.5 or less does not demonstrate quantum teleportation. The second limit corresponds to an idealized case of an infinite two-mode squeezing level ($\tilde{\nu}_{-}=0$), i.e., an EPR state, which realizes perfect quantum teleportation.

\begin{figure}
\includegraphics[width=0.48\textwidth]{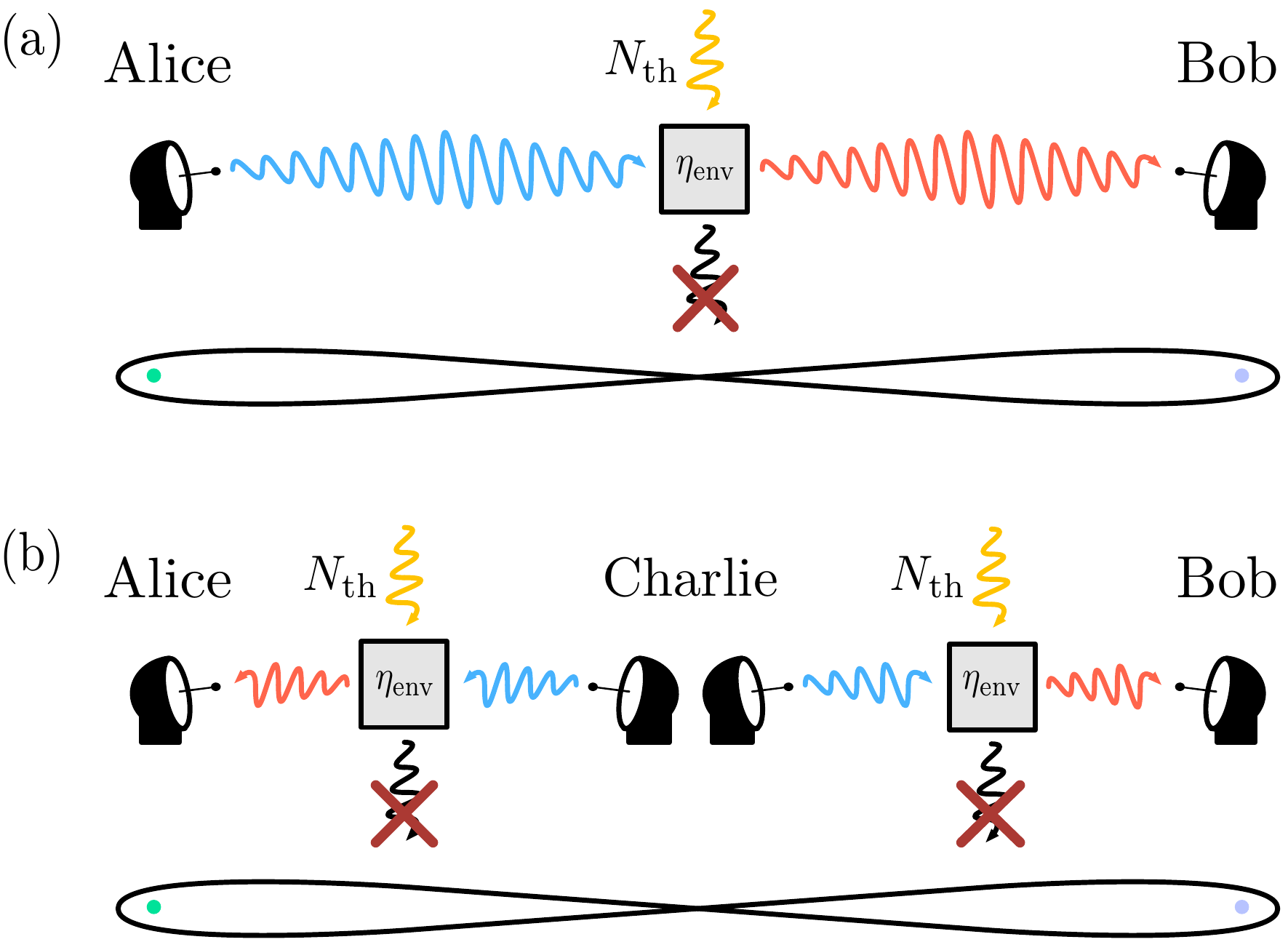}
\caption{Representation of an entanglement distribution protocol that uses antennae to efficiently transmit the quantum states into open-air, where photon losses and thermal noise effects are described with a beamsplitter with reflectivity $\eta_{\text{env}}$. We analyze two different scenarios: (a) Alice generates the entangled state, and attempts to share one of its modes with Bob by sending it through an noisy and lossy open-air channel that degrades the entanglement strength; (b) Charlie generates a two-mode entangled state, and sends one entngled mode to Alice and another to Bob. In this case, although both modes go through the same noisy and lossy channel, they travel half the distance as compared to the previous case.}
\label{fig2}
\end{figure}

\section{Open-air microwave entanglement distribution}\label{section:open_air_entanglement_distribution}
The scheme we have envisioned for open-air entanglement distribution relies on a variety of things, and its layout can be seen in Fig.~\ref{fig2}. First, generation of two-mode squeezed thermal (TMST) states with thermal photons $n < e^{-r}\sinh(r)$, given squeezing $r$. This will take place in a cryostat, at $ T \sim 50$ mK temperatures. Second, efficient transmission of states out of the cryostat and into open-air, targetting optimal entanglement preservation. For this task, we rely on an antenna based on the design proposed in Ref.~\cite{GonzalezRaya2020}, with a special attention to the shape of the impedance function, which greatly affects entanglement preservation. Third, estimation of losses in open-air which describe the attenuation of the signal caused by the presence of thermal noise in the environment, with the objective of setting bounds on effective transmission distances. This protocol will prepare the foundation for an open-air microwave quantum teleportation protocol, whose fidelity will depend on the entanglement of the resource shared by the parties involved.

\subsection{State Generation}\label{subsection:state_generation}                    
As the first step, we discuss the generation of entangled states inside a cryostat, more precisely, two-mode squeezed states. Squeezing is an operation in which one of the vriance of the electromagnetic field quadratures of a quantum state is reduced below the level of vacuum fluctuations, while the conjugate quadrature is amplified, satisfying the uncertainty principle. This can be achieved by sending the vacuum state to a JPA, a coplanar waveguide resonator line terminated by a direct current superconducting interference device (dc-SQUID). The dc-SQUID provides magnetic flux tunability to the resonator and enables parametric phase-sensitive amplification, which is the key for generating squeezed microwave states~\cite{Zhong2013,Pogorzalek2017}.

The relation between the frequency of the external magnetic flux, $\Omega$, and the fundamental frequency of the JPA, $\omega_{c}$, determines whether the JPA operates in the phase-insensitive or phase-sensitive regime. The latter is achieved in the so-called degenerate regime, $\Omega = 2\omega_{c}$. A corresponding three-wave mixing process, when one pump photon splits into two signal photons, is described by the Hamiltonian
\begin{equation}
H = g\left( \beta^{*}a^{2} - \beta a^{\dagger 2} \right).
\end{equation}
It can be shown that the aforementioned Hamiltonian corresponds to  a single-mode squeezing operator
\begin{equation}
S(\xi) = \exp\left[ \frac{1}{2}(\xi^{*}a^{2}-\xi a^{\dagger 2})\right],
\end{equation}
with the squeezing parameter given by $|\xi| \propto 2g|\beta|t$. 

A symmetric two-mode squeezed state can be generated by combining two, orthogonally-squeezed, states with equal squeezing levels at a hybrid ring. The latter element represents a symmetric 50:50 microwave beamsplitter. Microwave squeezed states produced by JPAs are subject to various sources of imperfections and noise. Therefore, the output states can be effectively modelled as two-mode squeezed thermal states, whose second moments differ from those of ideal two-mode squeezed vacuum states by a factor of $1+2n$, where $n$ is the number of thermal photons.

Thermal photons in squeezed states may have various physical origins. One of the most trivial reasons for noise in the two-mode squeezed states is finite temperatures of the input JPA modes, which leads to the fact that one applies squeezing operator to a thermal state rather than to a vacuum. Another important source of noise in squeezed states produced by flux-driven JPAs arises from Poisson photon number fluctuations in the pump mode, which leads to extra quasi-thermal photons in the output squeezed states~\cite{Renger2020}. Last but not least, higher-order nonlinear effects also contribute to additional effective noise under the Gaussian approximation~\cite{Boutin2017}.  More experimental details on the microwave squeezing and related imperfections can be found elsewhere~\cite{Fedorov2018}.


\subsection{Antenna Model \& Open-air losses}\label{subsection:antenna_model}

\begin{figure}
\includegraphics[width=0.48\textwidth]{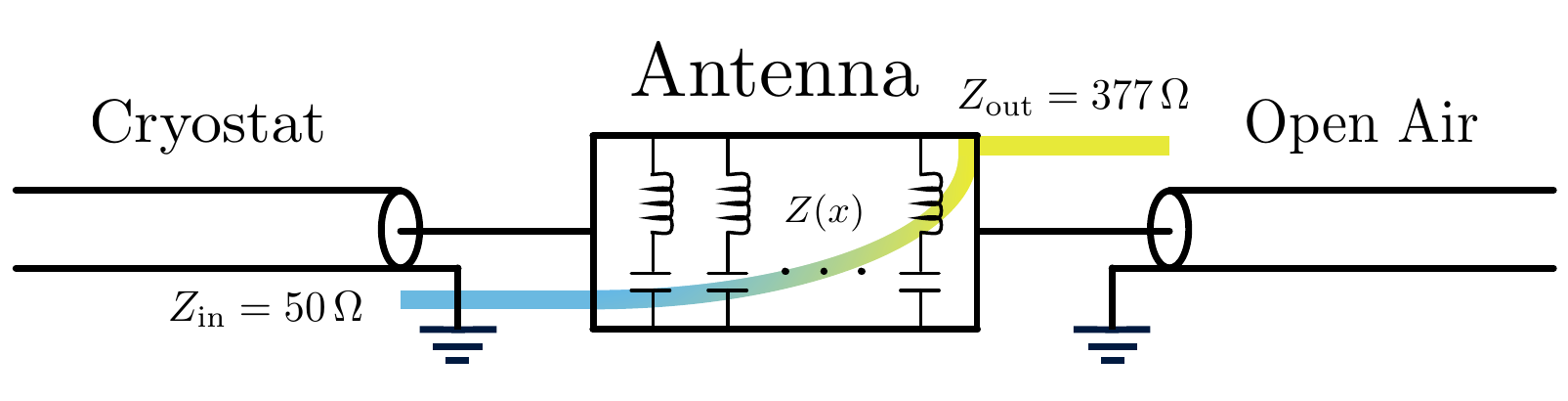}
\caption{Schematic representation of the circuit needed for a quantum communication protocol in which a quantum state is generated in a cryostat, and is then sent through an antenna into open-air, where transmission is assumed to be focused. As it is shown, a waveguide at impedance $Z_{\text{in}}=50 \, \Omega$ connects the cryostat with the antenna. The latter is a finite transmission line with variable impedance, $Z(x)$, designed to match the impedance of the cryostat with that of the open-air, $Z_{\text{out}}=377 \, \Omega$, maximizing the transmissivity. The final waveguide represents the directed propagation of the signal in open-air. }
\label{fig3}
\end{figure}

In transmission of quantum states from a cryostat into the open-air, an interface antenna comes into play as an inhomogeneous medium (as depicted in Fig.~\ref{fig3}) that connects those two very different environments. The main purpose of this antenna is to maximize transmission of the incident signal to the open-air medium. Here, we consider a transmission line coming out of the cryostat with an impedance of 50 $\Omega$, and assume focused transmission in open-air described by a transmission line with impedance 377 $\Omega$. Such antenna can be modelled by a finite transmission line with variable characteristic impedance and designed to match the impedances of the cryostat and of open-air at its ends. This approach was discussed in a previous study~\cite{GonzalezRaya2020}, where the transmitivity of the antenna was optimized for the task of entanglement distribution with TMST states of 5 GHz in an antenna of 3 cm. With an exponential profile of the impedance inside the antenna, reflectivity can be reduced down to $\sqrt{\eta}< 10^{-9}$, qualitatively matching the classical result of a horn antenna. 

Using this description, we can obtain a reflectivity coefficient that depends on experimental parameters, such as the length of the antenna, the carrier frequency of quantum states, and the internal and external impedances, among others. This result is compatible with the description of the antenna as a distributed beamsplitter, with its inputs being one of the modes of the TMST state together with a thermal state with $N_{\text{th}}$ thermal photons coming from the environment. From the two output modes, the reflected one is discarded, whereas the transmitted one is sent to Bob. The main antenna aim is to minimize the reflections of Alice's input signal in order to preserve entanglement between the transmitted mode and the retained one. 

Back-scattered thermal photons entering the emitting antenna might represent a certain problem for Alice and must be filtered. A straightforward, albeit somewhat challenging, solution for this problem is to use nonreciprocal microwave devices such as isolators (circulators) to protect entanglement-generating circuits from the unwanted thermal radiation.

\begin{figure}[t]
\vspace{0.5cm}
\includegraphics[width=0.48\textwidth]{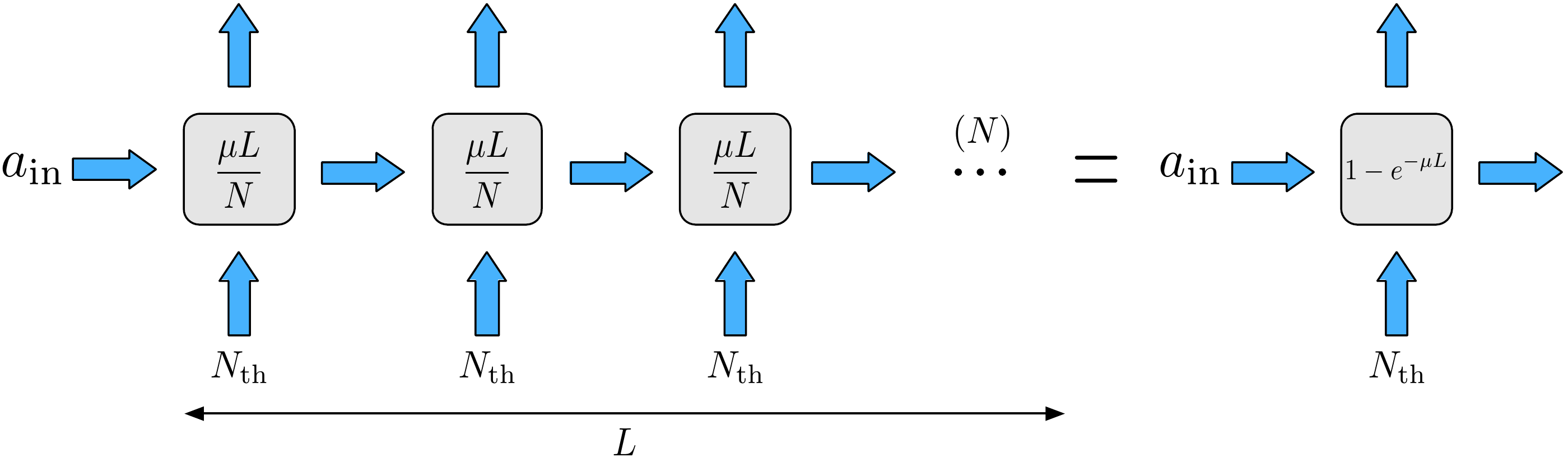}
\caption{Sketch of a beamsplitter loss model of an open-air quantum channel. Entanglement degradation of a state propagating in open-air at a constant temperature is modeled by an array of $N$ beamsplitters, each one introducing thermal noise characterized by $N_{\text{th}}$ thermal photons, assuming constant temperature throughout the path. An infinite array of beamsplitters ($N\rightarrow\infty$) can be approximated by a single beamsplitter with reflectivity $\eta_{\text{env}} = 1 - e^{-\mu L}$, where $L$ is the total channel length and $\mu$ is the reflectivity per unit length.}
\label{fig4}
\end{figure}

Once the state has been successfully sent out of the cryostat, we have to address the effects of entanglement degradation in open-air. Considering directed transmission in open-air, we envision an infinite array of beamsplitters to describe losses in open-air, as represented in Fig.~\ref{fig4}. Each one of these beamsplitters will allow for the mixing of thermal noise with the state. Assuming constant temperature throughout the sequence of possible absorption events, meaning that the thermal noise in each of the beamsplitters is characterized by $N_{\text{th}}$, we can obtain the reflectivity of an effective beamsplitter based on an attenutation channel~\cite{Serafini2017}, which represents the decay of quantum correlations and amplitudes,
\begin{equation}
\eta_{\text{env}} = 1-e^{-\mu L}.
\end{equation}
Here, $\mu$ represents a density of reflectivity, which in turn models photon losses per unit length, and $L$ is the travelled distance. This density of reflectivity can be interpreted as an attenuation coefficient, that quantifies the specific attenuation of signals in a given environment. In this work we consider $\mu = 1.44\cdot 10^{-6}\text{ m}^{-1}$ for the specific attenuation of 5 GHz signals caused by the presence of oxygen molecules in the environment (see Refs.~\cite{Ho2004,ITU-R}).

We could go further and assume that, attatched to the antenna (at constant temperature), there is another transmission line where the temperature is not constant throughout the trajectory, which leads to an inhomogeneous absorption probability. This is represented by density of reflectivity $\mu(x)$, and number of thermal photons $n(x)$. The latter still follows the Bose-Einstein distribution. An infinite array of beamsplitters that reproduce these features (see Ref.~\cite{GonzalezRaya2020}) can be replaced by a single beamsplitter with effective reflectivity and number of thermal photons given by
\begin{eqnarray}
\eta_{\text{env}} &=& 1-e^{-\int_{0}^{L} \diff x \mu(x)}, \\
\nonumber n_{\text{th}} &=& \frac{\int_{0}^{L} \diff x \mu(x)n(x) e^{-\int_{x}^{L}\diff x' \mu(x')}}{1-e^{-\int_{0}^{L} \diff x \mu(x)}},
\end{eqnarray}
where $L$ represents the total length of the array. Given that we are extending the length in which the transmission line remains at cryogenic temperatures, we will see that $n_{\text{th}} \leq N_{\text{th}}$.

Now that we have discussed how the signal is processed into the environment, let us characterize the resulting states. Assume that Alice generates a TMST state with $n$ thermal photons, and sends one mode to Bob over a distance $L$ through open-air, with a thermal background characterized by $N_{\text{th}}$ thermal photons. Then, the resulting state is what we call the `asymmetric' state:
\begin{widetext} 
\begin{equation}\label{CM_asym}
\Sigma_{\text{Asym}}=(1+2n)\begin{pmatrix} \left[\left(\frac{1+2N_{\text{th}}}{1+2n}\right) \eta_{\text{eff}} + (1-\eta_{\text{eff}})\cosh 2r\right]\Id_2 &  \sqrt{1-\eta_{\text{eff}}} \sinh 2r \sigma_Z  \\ 
\sqrt{1-\eta_{\text{eff}}} \sinh 2r \sigma_Z &  \cosh 2r \Id_2 \end{pmatrix},
\end{equation}
\end{widetext}
where $\eta_{\text{eff}}=1- e^{-\mu L}(1-\eta_\text{ant})$ represents the combined reflectivities of the antenna $\eta_{\text{ant}}$ and of the environment $\eta_{\text{env}}$. A sketch of the layout that leads to this kind of states can be seen in Fig.~\ref{fig2}~(a). With this, using Eq. \eqref{eq:ptseigen} we compute the partially-transposed symplectic eigenvalue,
\begin{equation}
\tilde{\nu}_{-}^{\text{out}} = \tilde{\nu}_{-}^{\text{in}} + \left( \frac{1}{2} + N_{\text{th}}\right)\eta_{\text{ant}}
\end{equation}
for very low reflectivities, $\eta_{\text{ant}} N_{\text{th}} \ll 1$, with $\tilde{\nu}_{\text{in}} = (1+2n)e^{-2r}$. See that, by reducing the reflectivity of the antenna, the impact of thermal noise is reduced, and the partially-transposed symplectic eigenvalue approaches that of the input state. In this extreme case, entanglement is fully preserved. 

Let us use the partially-transposed symplectic eigenvalue to compute the limit of entanglement. We will use the negativity as a measure of Gaussian entanglement, $\mathcal{N} = \max\{ 0, \frac{1-\tilde{\nu}_{-}}{2\tilde{\nu}_{-}}\}$, such that this limit occurs for $\tilde{\nu}_{-}=1$. This constitutes a bound on the reflectivity; all smaller values of $\eta_{\text{eff}}$ will result in entanglement preservation. This result is
\begin{equation}\label{ref_bound}
\eta_\text{max} = \frac{1}{1+\frac{N_{\text{th}}}{1+\frac{2n(1+n)}{1-(1+2n)\cosh(2r)}}},
\vspace{0.15cm}
\end{equation}
together with the conditions $n < e^{-r}\sinh(r)$ and $r > 0$. With this bound, we can obtain the maximum distance entanglement can survive,
\begin{equation}
L_\text{max} = -\frac{1}{\mu}\log(1-\eta_\text{max}).
\end{equation} 
Imagine that TMST states are generated in the cryostat at $50$ mK temperature, with thermal photons $n \sim 10^{-2}$,  and squeezing $r = 1$. In open-air, at $300$ K, the number of thermal photons is $N_{\text{th}} \sim 1250$. Assuming a perfect antenna ($\eta_{\text{ant}}=0$), the maximum distance the state can travel before entanglement completely degrades is $L_\text{max} \sim 550$ m.


As a different approach to the entangled resource, we assume that a TMST state is generated at an intermediate spot between both parties, and that each mode is sent through an antenna and travels some distance $L_i$, with $i=\{1,2\}$, before reaching Alice and Bob. Then, each mode will see an effective reflectivity of $\eta^{(i)}_{\text{eff}}= 1 - e^{-\mu L_i}(1-\eta_\text{ant})$, combining the effects of the antenna and the environment. We assume for simplicity that $L_1+L_2 = L$, where $L$ is the linear distance between Alice and Bob. The covariance matrix of such a state, which we refer to as `symmetric', is
\begin{widetext} 
\begin{equation}\label{CM_sym}
\Sigma_{\text{Sym}} =(1+2n)\begin{pmatrix} \left[\left(\frac{1+2N_{\text{th}}}{1+2n}\right)\eta^{(1)}_{\text{eff}}  + (1-\eta^{(1)}_{\text{eff}})\cosh 2r\right]\Id_2 & \sqrt{\left(1-\eta^{(1)}_{\text{eff}}\right)\left(1-\eta^{(2)}_{\text{eff}}\right)} \sinh 2r \sigma_Z  \\ 
\sqrt{\left(1-\eta^{(1)}_{\text{eff}}\right)\left(1-\eta^{(2)}_{\text{eff}}\right)} \sinh 2r \sigma_Z & \left[ \left(\frac{1+2N_{\text{th}}}{1+2n}\right)\eta^{(2)}_{\text{eff}}  + (1-\eta^{(2)}_{\text{eff}})\cosh 2r\right]\Id_2 \end{pmatrix},
\end{equation}
\end{widetext}
and corresponds to the layout represented in Fig.~\ref{fig2}~(b). With this state, the maximum distance entanglement can survive is $L_\text{max} \sim 480$ m.

Throughout this manuscript, we will refer to these states as the (asymmetric and/or symmetric) lossy TMST states, the bare states, or the TMST states distributed through open-air. 

Furthermore, we could also consider the specific attenuation caused by the presence of water vapor in the environment~\cite{Ho2004}. This would lead to higher attenuation coefficients, thus reducing the distances that entanglement can survive. For an average water vapor density, these distances are 450 m and 390 m for asymmetric and symmetric states, respectively. They become 400 m for asymmetric states and 350 m for symmetric states in a maximum water vapor density scenario.


\subsection{Overcoming entanglement degradation: distillation and swapping with microwaves}\label{subsection:distillation_and_swapping}
We have seen that entanglement distribution between two parties is limited by environmental noise, as well as by photon losses. Considering we have a perfect antenna, these factors can limit the maximum distance entanglement can survive to a few hundred meters. Since an amplification protocol only contributes to the degradation of quantum correlations (see Appendix~\ref{app_A}), we investigate entanglement distillation and entanglement swapping, two techniques which could improve both the reach and the quality of entanglement, at the expense of efficiency~\cite{DiCandia2015}. 

\subsubsection{Entanglement distillation}
This technique aims at increasing entanglement in quantum states by means of local operations. Let us briefly review different ways to distill entanglement. One of them is noiseless linear amplification, a non-deterministic operation~\cite{Ralph2009,Xiang2010} that requires unincreasing distinguishability of amplified states, and which has been recently achieved in the microwave regime~\cite{Dassonneville2020}. It also requires efficient photocounting, which is where the non-deterministic part comes into play. At the core of this protocol lies a process based on the quantum scissors~\cite{Pegg1998}. The gain of this procedure is inversely proportional to the success probability, which also decreases with increasing number of resources, making it very costly. 

Another widely-known protocol is Gaussian distillation, which is also non-deterministic, but it requires only two initial copies of a state, as well as efficient photodetection. If the incoming entangled state is Gaussian, then it is initially de-Gaussified by combining two copies of said state with balanced beamsplitters and keeping the transmitted state when any number of photons has been detected at the reflected modes~\cite{Browne2003}. Another possible de-Gaussification protocol applies an operation $\hat{Z}=(1-\omega)a^{\dagger}a+\omega a a^{\dagger}$~\cite{Fiurasek2010} on a quantum state without requiring a copy. Gaussian distillation begins when two copies of the resulting state are mixed by $50:50$ beamsplitters and, if no photons are reflected, the operation is applied again~\cite{Browne2003,Eisert2004}. Provided that the initial states were entangled, this process leads to a non-Gaussian state with higher entanglement. However, it is also costly in terms of the number of resources, and it only produces a state that is Gaussian (and with higher entanglement) in an infinite-application limit of the Gaussification channel.  

\begin{figure}
\includegraphics[width=0.45\textwidth]{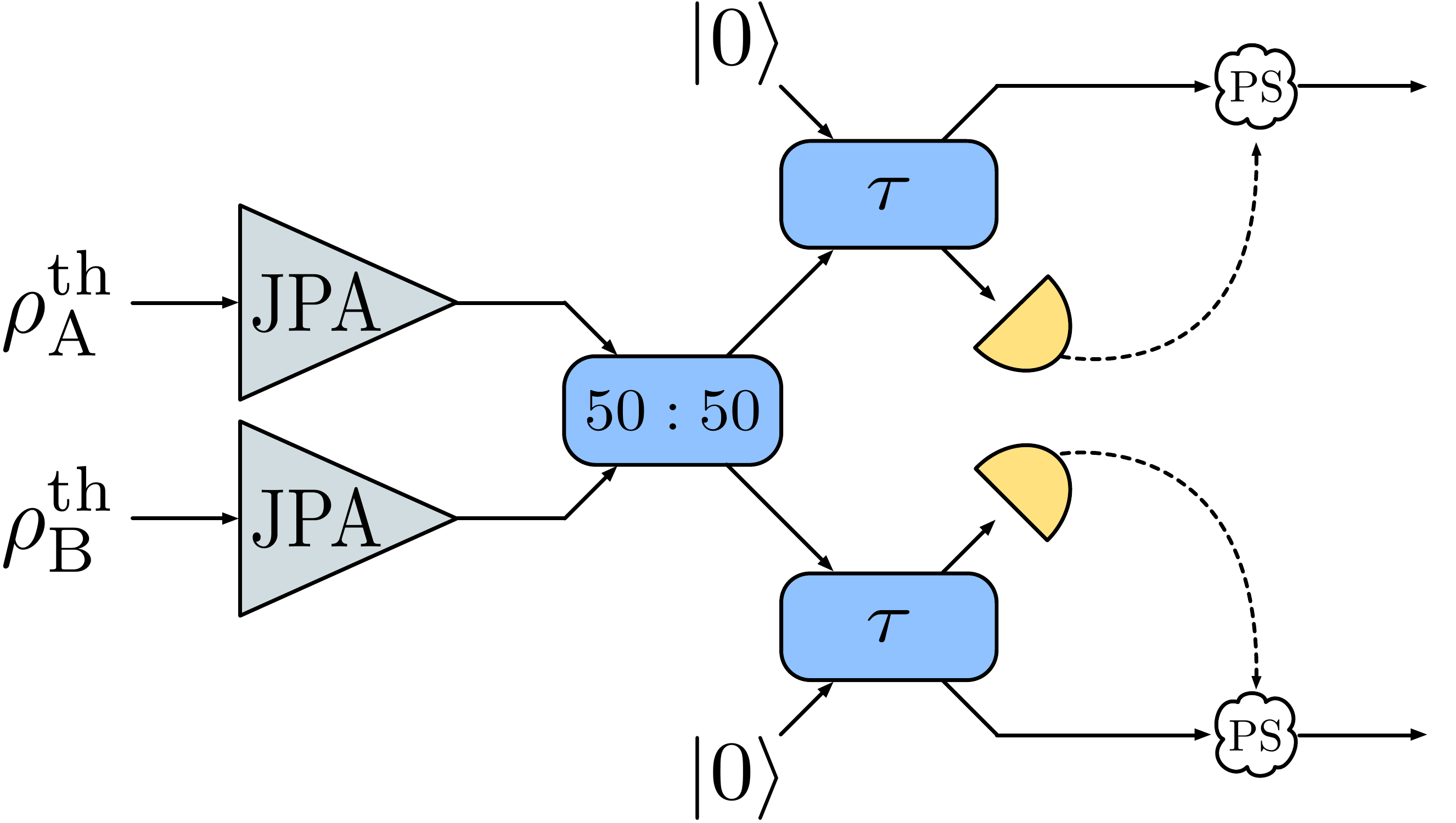}
\caption{Sketch of a photon subtraction scheme applied to two-mode squeezed thermal states, which are generated from single-mode squeezed thermal states by JPAs, subsequently combined in a balanced beamsplitter. Each mode of the resulting state is then combined with an ancillary vacuum state in high-transmissivity (we consider $\tau = 0.95$) beamsplitters, with photocounters placed at each reflected path. The resulting state shows higher entanglement for low values of the squeezing parameter, where the limit for enhancement will vary depending on the number of photons detected.}
\label{fig5}
\end{figure}

Finally, we review another non-deterministic protocol, which does not require the storage or production of simultaneous copies of a quantum state, and whose gain is also inversely proportional to the success probability. This protocol is called probabilistic photon subtraction~\cite{Takahashi2010}, and it utilizes non-Gaussian operations in order to distill entanglement, as we have seen in the previous protocols. However, in this situation, we will not look to re-Gaussify the state afterwards. A discussion about the heuristic photon subtraction protocol with TMSV states, the more theoretical approach that does not consider the effects of beamsplitters and measurement, can be found in Appendix~\ref{app_B}.

Probabilistic photon subtraction starts with an entangled state, for example a two-mode squeezed vacuum (TMSV) state. This can be produced by two single-mode squeezed states with squeezing parameter $r$, which are combined by a $50:50$ beamsplitter, as shown in Fig.~\ref{fig5}, resulting in a TMSV state,
\begin{equation}
\sqrt{1-\lambda^{2}}\sum_{n=0}^{\infty} \lambda^{n}|n,n\rangle_{AB}
\end{equation}
with $\lambda = \tanh(r)$. The next step of the protocol is to mix each mode with an ancillary vacuum state at two highly-transmitting, identical beamsplitters. The output photon-subtracted state is postselected depending on the outcome of the photocounts performed at each beamsplitter. Here, we shall focus on photon-subtracted TMSV states where the same number of photons are subtracted from each mode. The resulting $2k$-photon subtracted TMSV state is then:
\begin{equation}\label{eq:kPSstate}
|\psi^{(2k)}\rangle_{AB} = P_{2k}^{-1/2}\sum_{n=0}^{\infty} a^{(k)}_n\ket{n,n}_{AB},
\end{equation}
with $a^{(k)}_n \equiv  \sqrt{1-\lambda^{2}}\lambda^{n+k}(-1)^{2k} \begin{pmatrix}n+k\\k\end{pmatrix} (1-\tau)^{k} \tau^{n}$, and $P_{2k} \equiv \sum_{n=0}^\infty \abs{a^{(k)}_n}^2$, which can be interpreted as the probability of successfully subtracting $k$ photons from each mode of a TMSV state. The sum converges to $P_{2k} = \left(1-\lambda ^2\right) (\lambda -\lambda_\tau)^{2 k} \, _2F_1\left(k+1,k+1;1;\lambda_\tau ^2\right)$ where $\, _2F_1\left(a,b;c;z\right)$ is the Gaussian hypergeometric function, and $\lambda_\tau\equiv \tau \lambda$.
In what follows, we will focus on the cases $k=1,2$, which correspond to two-photon subtraction (2PS) and four-photon subtraction (4PS), respectively, and whose corresponding success probabilities are
\begin{eqnarray}\label{eq:success-prob-TMSV}
\nonumber P_{2} &=& (1-\lambda^{2})\lambda^{2}(1-\tau)^{2}\frac{(1+\lambda_{\tau}^{2})}{(1-\lambda_{\tau}^{2})^{3}}, \\
P_{4} &=& 4\left(1-\lambda ^2\right) \lambda ^4  (1-\tau)^4\frac{\left(1+\lambda_{\tau}^4+4 \lambda_{\tau} ^2\right)}{\left(1-\lambda_{\tau}^2\right)^5}.
\end{eqnarray}

\begin{figure}
\centering
\includegraphics[width=0.48\textwidth]{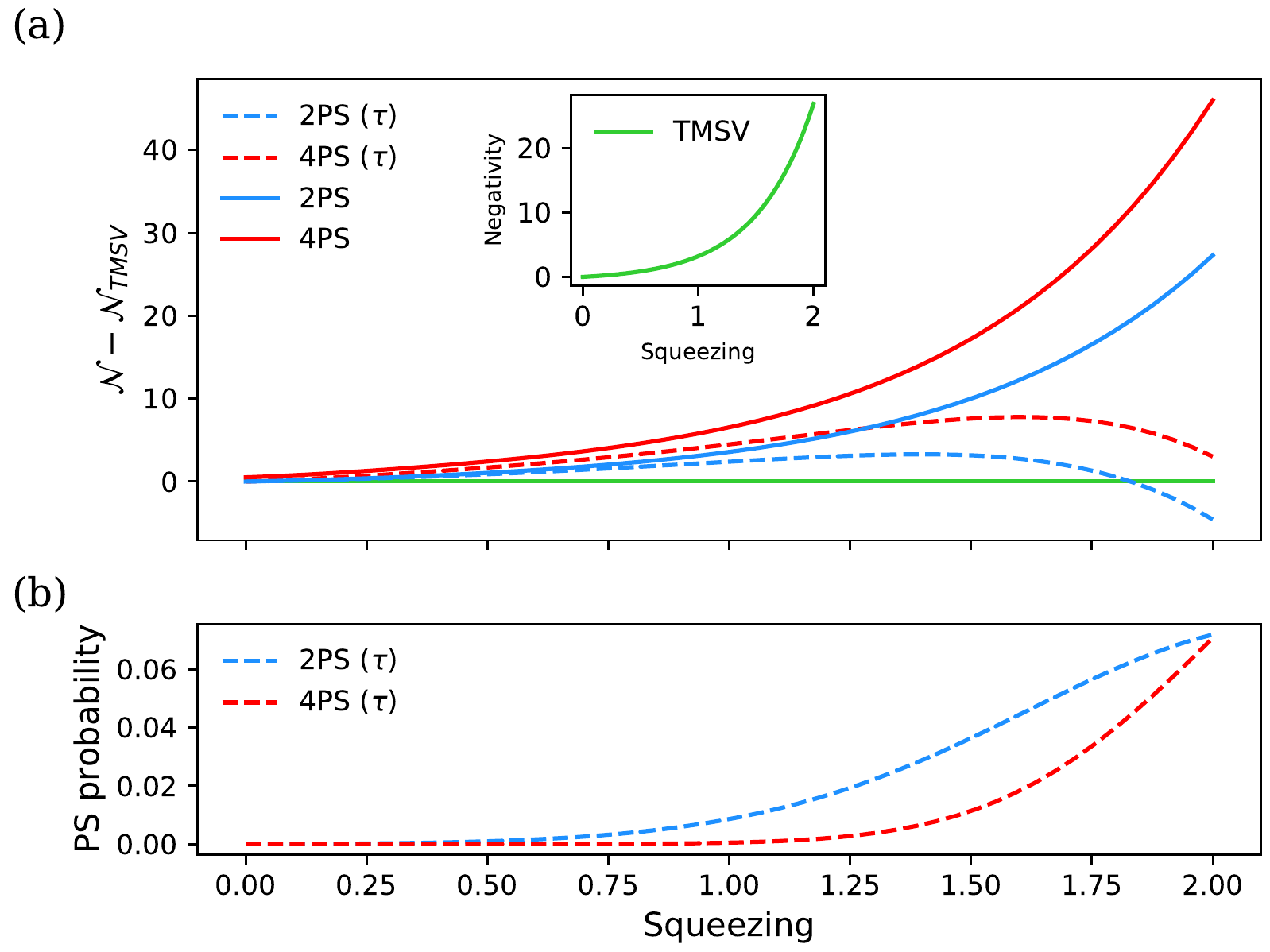}
\caption{(a) Negativity difference between photon-subtracted and bare two-mode squeezed vacuum (TMSV) states, represented against the initial squeezing parameter. The blue and red curves represent two-and-four photon subtraction, respectively, whereas the green curve represents the no-gain line, above which any point represents an improvement in negativity. Curves associated to probabilistic photon subtraction appear dashed, whereas the solid ones are associated to heuristic photon subtraction. We have considered the transmissivity of the beamsplitters involved in probabilistic photon subtraction to be $\tau=0.95$. As an inset, we represent the negativity curve for the TMSV state versus initial squeezing parameter. (b) Sucess probability of symmetric photon subtraction schemes: two-photon subtraction (2PS) is displayed in a blue-dashed line, and four-photon subtraction (4PS) is represented in a red-dashed line.}
\label{fig6}
\end{figure}

If photon subtraction is successful for any (non-zero) number of photons, the resulting state shows increased entanglement with respect to the TMSV in a certain interval. This can be seen by computing the negativity $\mathcal{N}(\rho^{(2k)})$ of the family of states \eqref{eq:kPSstate}. We find that for $\rho^{(2k)} \equiv \ket{\psi^{(2k)}}\bra{\psi^{(2k)}}$ the negativity is
\begin{equation}
\mathcal{N}(\rho^{(2k)}) =  \frac{A_k-1}{2},
\end{equation}
where \begin{equation}
A_k \equiv \frac{\left(\sum_{n=0}^\infty a^{(k)}_n\right)^2}{P_{2k}}.
\end{equation}
Performing the sum we obtain
\begin{equation}
\mathcal{N}(\rho^{(2k)})=\frac{1}{2} \left(\frac{(1- \lambda_\tau)^{-2 (k+1)}}{\, _2F_1\left(k+1,k+1;1; \lambda_\tau^2\right)}-1\right),
\end{equation}
which describes the negativity of the heuristic photon subtraction protocol (see Appendix~\ref{app_B}) in the limit $\tau\rightarrow 1$, while reproducing the negativity of the TMSV state, $\mathcal{N}_{\text{TMSV}}=\lambda/(1-\lambda)$, in the case $\tau\rightarrow1$ and $k=0$.

In Fig.~\ref{fig6}~(a) we represent negativity differences as a function of the initial squeezing $r$. We subtract the negativity of the TMSV state to those of the 2PS (blue) and the 4PS (red), heuristic (solid), and probabilistic (dashed), with a beamsplitter transmissivity $\tau = 0.95$. See that probabilistic photon subtraction works for lower squeezing, while heuristic photon subtraction is always advantageous. In Fig.~\ref{fig6}~(b), we display the success probability of two-photon (blue, dashed) and four-photon (red, dashed) subtraction. Observe that 2PS shows higher probability than 4PS, whereas the latter shows higher improvement than the former. As the squeezing parameter increases, both probabilities grow closer, as probabilistic PS losses its advantage.

The rate of two-mode squeezed state generation is defined by the effective bandwidth of JPAs. In the case of conventional resonator-based JPAs, these bandwidths are typically on the order of $\sim 10$\, MHz~\cite{Pogorzalek2017}. By exploiting more advanced designs based on travelling-wave Josephson parametric amplifiers, one can hope to increase these bandwidth to $\sim 1$\,GHz. However, the price for this increase is typically lower squeezing levels and higher noise photon numbers.

When dealing with our lossy TMST states, we consider a two-photon subtraction protocol which is applied right before the teleportation experiment, in order to prepare the entangled resource for an enhanced performance. That means, we apply photon subtraction on states that have already travelled through open-air. In the case of TMSV states, photon subtraction was beneficial for low squeezing, which translates into low entanglement. In the case of lossy TMST states, entanglement is affected by the initial squeezing, but also by the distance, since reducing it means reducing photon losses and the presence of thermal noise in the state. Consequently, a better performance of the teleportation protocol using photon-subtracted entangled states as the resource occurs for small distances. The submatrices of the covariance matrix that characterizes the entangled resource, written as $\Sigma_{A} = \alpha\mathbb{1}_{2}$, $\Sigma_{B} = \beta\mathbb{1}_{2}$, and $\varepsilon_{AB} = \gamma\sigma_{Z}$, are modified by a symmetric two-photon subtraction process as follows
\begin{widetext}
\begin{eqnarray}\label{2PS_submatrices}
\nonumber \tilde{\Sigma}_{A} &=& \left[1-2\tau\frac{(1-\alpha)(1+\beta)+\gamma^{2}+((1-\alpha)(1-\beta)-\gamma^{2})\tau}{(1+\alpha)(1+\beta)-\gamma^{2}+2(1-\alpha\beta+\gamma^{2})\tau+((1-\alpha)(1-\beta)-\gamma^{2})\tau^{2}}\right]\mathbb{1}_{2}, \\
\nonumber \tilde{\Sigma}_{B} &=& \left[1-2\tau\frac{(1+\alpha)(1-\beta)+\gamma^{2}+((1-\alpha)(1-\beta)-\gamma^{2})\tau}{(1+\alpha)(1+\beta)-\gamma^{2}+2(1-\alpha\beta+\gamma^{2})\tau+((1-\alpha)(1-\beta)-\gamma^{2})\tau^{2}}\right]\mathbb{1}_{2}, \\
\tilde{\varepsilon}_{AB} &=& \frac{4\tau\gamma}{(1+\alpha)(1+\beta)-\gamma^{2}+2(1-\alpha\beta+\gamma^{2})\tau+((1-\alpha)(1-\beta)-\gamma^{2})\tau^{2}}\sigma_{Z},
\end{eqnarray}
whose success probability is given by
\begin{equation}\label{eq:success-prob}
P = 4(1-\tau)^{2}\frac{\left[1-\alpha\beta+\gamma^{2}+((1-\alpha)(1-\beta)-\gamma^{2})\tau\right]^{2}-(\alpha-\beta)^{2}+4\gamma^{2}}{\left[(1+\alpha)(1+\beta)-\gamma^{2}+2(1-\alpha\beta+\gamma^{2})\tau+((1-\alpha)(1-\beta)-\gamma^{2})\tau^{2}\right]^{3}}. 
\end{equation}
\end{widetext}
See Appendix~\ref{app_C} for the general expressions. In order for these submatrices to characterize a covariance matrix, they need to satisfy a positivity condition, as well as the uncertainty principle. Both these requirements can be summarized by one,
\begin{equation}
\left|\sqrt{\det\tilde{\Sigma}} - 1 \right| \geq \left| \tilde{\alpha}-\tilde{\beta}\right|,
\end{equation}
given that we have used $\tilde{\Sigma}_{A} = \tilde{\alpha}\mathbb{1}_{2}$, $\tilde{\Sigma}_{B} = \tilde{\beta}\mathbb{1}_{2}$. Unfortunately, the state resulting from photon subtraction is not Gaussian, and then, it is not completely characterized by the covariance matrix. In this case, we use the characteristic function to describe the photon-subtracted TMST states. The general expression for the characteristic function of a probabilistically two-photon-subtracted Gaussian state (where one photon has been subtracted in each mode) can be found in Appendix~\ref{app_C}. Furthermore, please see Appendix~\ref{app_B} for an equivalent discussion regarding heuristic photon subtraction.

\subsubsection{Entanglement swapping}
In this section we contemplate the CV version of entanglement swapping~\cite{Hoelscher-Obermaier2011}, a procedure that can be used to reduce the distance which states have to travel through the environment, and hence attenuate the effects of entanglement degradation. We consider the case in which we have two entangled states, shared by three parties pairwisely. That is, between Alice and Charlie, and between Charlie and Bob. Entanglement swapping is a technique that allows for the conversion of two bipartite entangled states into a single one shared by initially-unconnected parties. By making measurements in a maximally-entangled basis, Charlie is able to transform the entangled resources he shares with Alice and with Bob into a single entangled state shared only by Alice and Bob. In CV, these measurements are described by Homodyne detection, and their effect on the state is computed as we have seen in the CV teleportation protocol. Consider that these states are Gaussian, with covariance matrices
\begin{eqnarray}
\nonumber \Sigma_{1} &=& \begin{pmatrix} \Sigma_{A} & \varepsilon_{AB} \\ \varepsilon_{AB}^{\intercal} & \Sigma_{B} \end{pmatrix}, \\
\Sigma_{2} &=& \begin{pmatrix} \Sigma_{C} & \varepsilon_{CD} \\ \varepsilon_{CD}^{\intercal} & \Sigma_{D} \end{pmatrix},
\end{eqnarray}
and null displacement vectors. Then, the covariance matrix of the remaining state,
\begin{equation}
\Sigma^{\text{ES}} = \begin{pmatrix} \tilde{\Sigma}_{A} & \tilde{\varepsilon}_{AD} \\ \tilde{\varepsilon}_{AD}^{\intercal} & \tilde{\Sigma}_{D} \end{pmatrix},
\end{equation}
conditioned by the measurement results is characterized by
\begin{eqnarray}
\nonumber \tilde{\Sigma}_{A} &=& \Sigma_{A} - \frac{\varepsilon_{AB}\left(\Sigma_{B}+\sigma_{Z}\Sigma_{C}\sigma_{Z}\right)\varepsilon_{AB}^{\intercal}}{\det\left(\Sigma_{B}+\sigma_{Z}\Sigma_{C}\sigma_{Z}\right)}, \\
\nonumber \tilde{\Sigma}_{D} &=& \Sigma_{D} - \frac{\varepsilon_{CD}\left(\Sigma_{C}+\sigma_{Z}\Sigma_{B}\sigma_{Z}\right)\varepsilon_{CD}^{\intercal}}{\det\left(\Sigma_{B}+\sigma_{Z}\Sigma_{C}\sigma_{Z}\right)}, \\
\tilde{\varepsilon}_{AD} &=& \frac{\varepsilon_{AB}\left(\Sigma_{B}\sigma_{Z}+\sigma_{Z}\Sigma_{C}\right)\varepsilon_{CD}^{\intercal}}{\det\left(\Sigma_{B}+\sigma_{Z}\Sigma_{C}\sigma_{Z}\right)}.
\end{eqnarray}
We have observed that, in the setup we are considering, the only protocol that presents an improvement in negativity with respect to the bare states is the one in which Alice and Bob generate the two-mode entangled states, and each send one of the modes to Charlie. Then, the two modes used for entanglement swapping are the ones that have become mixed with environmental noise. Nevertheless, this enhancement occurs for large distances, which implies low negativities, and works significantly better in low-temperature environments, where $N_{\text{th}}$ is reduced. Considering $\Sigma_{A} = \Sigma_{D} = \alpha\mathbb{1}_{2}$, $\Sigma_{B} = \Sigma_{C} = \beta\mathbb{1}_{2}$, and $\varepsilon_{AB} = \varepsilon_{CD} = \gamma\sigma_{Z}$, then we can characterize the covariance matrix of the resulting resource by
\begin{eqnarray}\label{ES_submatrices}
\nonumber \tilde{\Sigma}_{A} &=& \left( \alpha - \frac{\gamma^{2}}{2\beta}\right)\mathbb{1}_{2}, \\
\nonumber \tilde{\Sigma}_{D} &=& \left( \alpha - \frac{\gamma^{2}}{2\beta}\right)\mathbb{1}_{2}, \\
\tilde{\varepsilon}_{AD} &=& \frac{\gamma^{2}}{2\beta}\sigma_{Z}.
\end{eqnarray}
The condition for this characterization to be appropriate is given by
\begin{equation}
\left|\sqrt{\det\Sigma_{1}}-\frac{\beta}{\alpha}\right| \geq 0.
\end{equation}
Please, see Appendix~\ref{app_D} for further discussion. 

\begin{figure}
\centering
\includegraphics[width=0.48\textwidth]{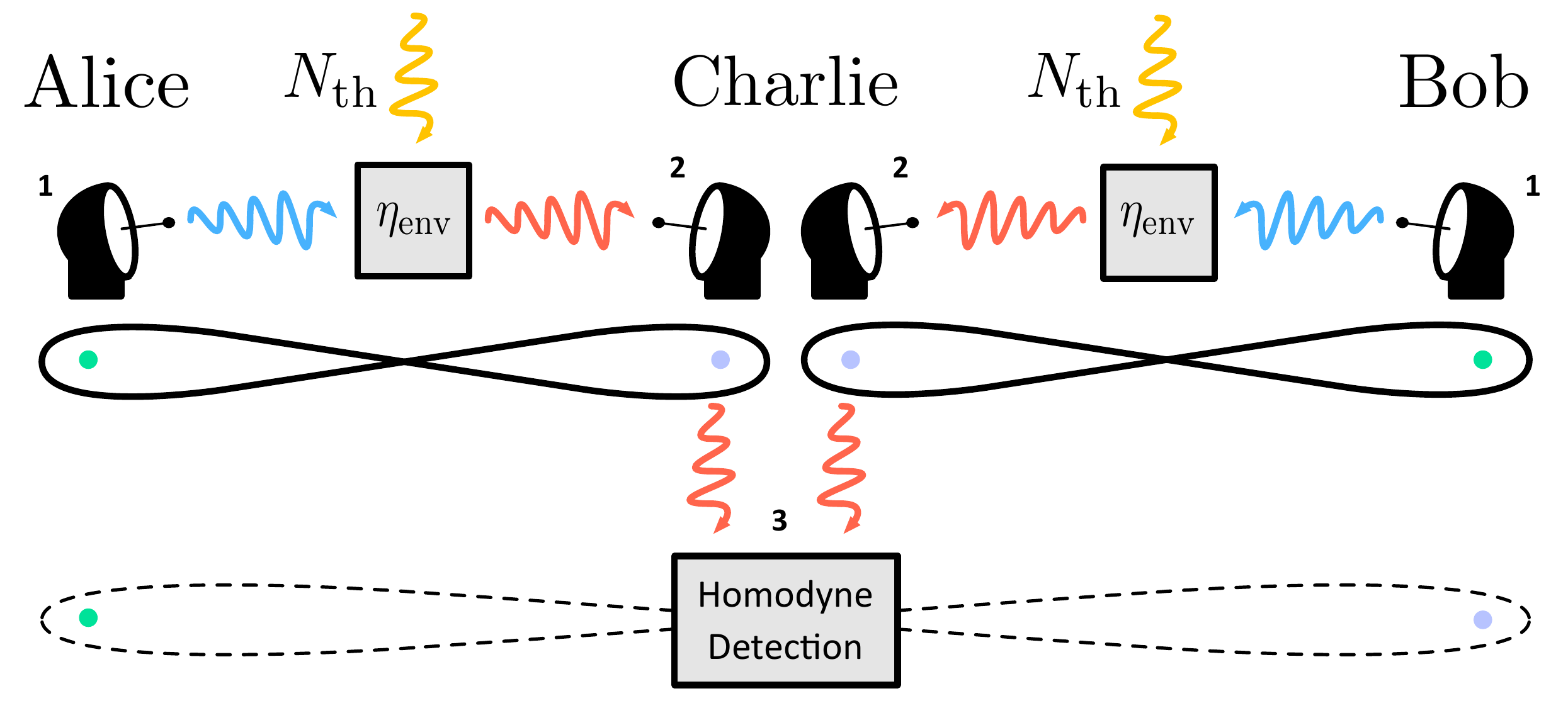}
\caption{Sketch of the optimal entanglement swapping scheme involving three parties, and three key steps: first, Alice and Bob generate two-mode squeezed thermal states and, while keeping one of the modes each, send the others through open-air, where they are subject to photon loss and get mixed with thermal noise. Second, Charlie receives and processes both modes, and third, he uses them to perform Homodyne Detection. In the end, Charlie is able to transform the pairwisely-entangled states he shares with Alice and with Bob independently into an entangled state held solely between Alice and Bob.}
\label{fig7}
\end{figure}


\section{Experimental limitations to photocounting and homodyning with microwaves}\label{section:homodyning_and_photocounting}
In this section, we review current advances on photocounting and homodyne detection techniques with microwave quantum technologies. These techniques are vital for photon subtraction, as well as for entanglement swapping and quantum teleportation, which are the processes described in this manuscript.

\subsection{Photodetection}
traditionally, the problem of detecting microwaves has been the low energy of the signals when compared to the optical regime. Any of the entanglement distillation protocols we have discussed will require some kind of photodetection scheme. Particularly, for photon subtraction, a photocounter for microwave photons is required. In the current landscape of microwave quantum technologies, there have been recent proposals for non-demolition detection of itinerant single microwave photons~\cite{Kono2018,Besse2018,Lescanne2020} in circuit-QED setups, with detection efficiencies ranging from $58\%$ to $84\%$. Based on similar setups, a photocounter has been proposed~\cite{Dassonneville2020} that can detect up to 3 microwave photons. 

This device is able to catch an incoming wavepacket into a buffer resonator, which is then transferred into the memory by means of pumping a Josephson ring modulator. Then, the information about the number of photons in the memory is transferred to a transmon qubit, which is coupled to the memory modes, and from where it is read bit by bit. Consequently, this photocounter requires previous knowledge on the waveform and the arrival time of the incoming mode to be detected. Furthermore, this device is not characterized by a single quantum efficiency; rather, the detection eficiency varies depending on the number of photons. That is, $99\%$ for zero photons, $(76\pm3)\%$ for a single photon, $(71\pm3)\%$ for two photons and $(54\pm2)\%$ for three, assuming a dark count probability of $(3\pm0.2)\%$ and a dead time of $4.5\,\mu$s.


\subsection{Homodyne detection}
Homodyne detection allows one to extract information about a single quadrature. It can be used to perform CV-Bell measurements, i.e. a projective measurement in a maximally-quadrature-entangled basis for CV states. One way to perform Bell measurements with propagating CV states is to use the analog feedforward technique, as demonstrated in Ref.~\cite{Fedorov2021}. This approach requires operating two additional phase-sensitive amplifiers in combination with two hybrid rings and a directional coupler, which effectively implements a projection operation for conjugate quadratures of propagating electromagnetic fields. An alternative, more conventional approach can be implemented by using novel microwave single-photon detectors and adapting them to the well-known optics homodyning techniques.

As we have seen, entanglement swapping provides an advantage if this measurement scheme is used without averaging over the results (single-shot homodyning), whereas the Braunstein-Kimble quantum teleportation protocol assumes this average is performed, given an unknown coherent state. In theory, single-shot homodyning can be implemented by using quantum-limited superconducting amplifiers and standard demodulation techniques~\cite{Eichler2012}. However, some fundamental aspects of the ``projectiveness'' of this operation and its importance for the Bell detection measurements or for photon subtraction are still unclear and must be verified.

\section{Open-Air Microwave Quantum Teleportation Fidelities}\label{section:teleportation}

In this section, we compute the average teleportation fidelity for different resource states. In all cases, the teleported state is a coherent state $\ket{\alpha_0}\bra{\alpha_0}$. 

\subsection{Two-mode squeezed vacuum resource}
The case of a TMSV state is particularly simple, as we can simply plug its covariance matrix into Eq.~\eqref{eq:Gaussian-fidelity},
\begin{equation}
\overline{F}_{{\text{TMSV}}} = \frac{1+\lambda}{2}.
\end{equation}
When symmetric $2k$-photon subtraction is performed, the formula for Gaussian average fidelity can no longer be invoked. The results for $k=1,2$ (2PS and 4PS respectively) are:
\begin{eqnarray}
\overline{F}_{\text{2PS}} &=& \left(1-\lambda_{\tau} + \frac{\lambda_{\tau}^{2}}{2}\right)\frac{(1+\lambda_{\tau})^{3}}{2(1+\lambda_{\tau}^{2})}, \\
\nonumber \overline{F}_{\text{4PS}} &=& \frac{(1+\lambda_{\tau} )^{5}\left[ 8-\lambda_{\tau}(2-\lambda_{\tau})(8-3\lambda_{\tau}(2-\lambda_{\tau})) \right]}{16(1+4\lambda_{\tau}^2+\lambda_{\tau}^4)},
\end{eqnarray}
with $\lambda_{\tau}=\lambda\tau$.
\begin{figure}
\includegraphics[width=0.5 \textwidth]{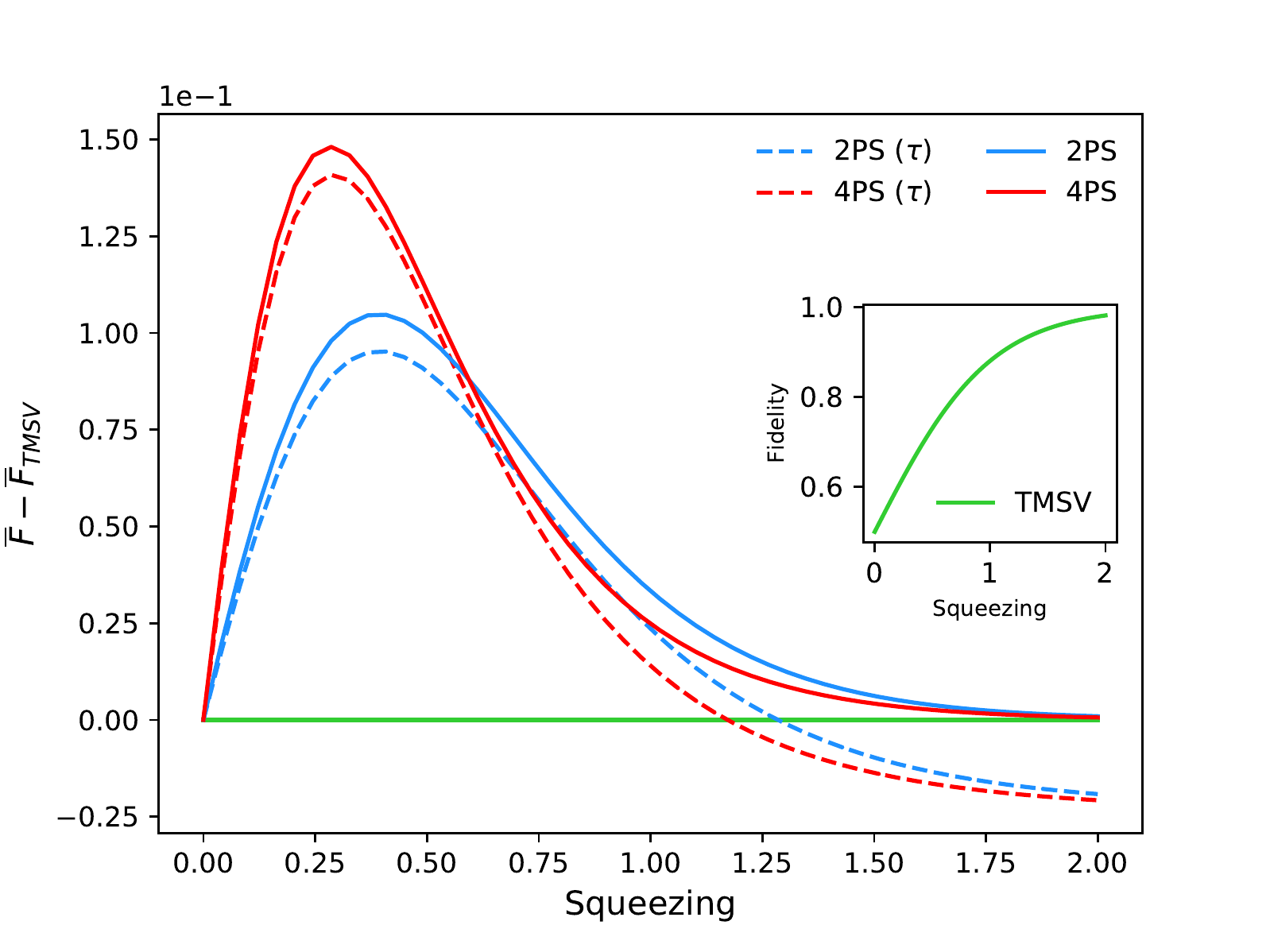}
\caption{Average fidelity of CV quantum teleportation of an unknown coherent state with respect to the initial squeezing parameter. We subtract the average fidelity associated with a two-mode squeezed vacuum (TMSV) state resource to the average fidelities of two-photon subtracted (2PS, blue) and four-photon subtracted (4PS, red) TMSV states. Curves associated to probabilistic photon subtraction appear dashed, whereas the solid ones are associated to heuristic photon subtraction. The green curve describes the TMSV case, which delimits the no-gain line, above which any point represents an improvement in fidelity due to photon subtraction. As an inset, we plot the average fidelity associated to a TMSV state against the initial squeezing parameter. We have considered the transmissivity of the beamsplitters involved in probabilistic photon subtraction to be $\tau=0.95$.}
\label{fig8}
\end{figure}

In Fig.~\ref{fig8}, we represent the result of subtracting the fidelity associated to the bare TMSV state to those associated to two-photon (2PS, blue) and four-photon (4PS, red) subtracted TMSV states. Fidelity differences associated to heuristic photon subtraction appear as solid lines, whereas those associated to probabilistic photon subtraction appear dashed. The green solid line represents the no-gain line, above which any photon-subtracted state presents an advantage in fidelity. Notice that photon subtraction works better for low squeezing, and as we increase it we see that using the TMSV state as a resource for teleportation renders a higher fidelity than probabilistic photon subtraction, while heusristic photon subtraction tends to the TMSV result. 

\subsection{Two-mode squeezed thermal resource}
We now study the teleportation fidelity associated to a two-mode squeezed thermal state, sent through a lossy and noisy channel defined by the combination of the antenna and an environment with $N_\text{th}$ photons. By defining $\Sigma_{A}=\alpha\mathbb{1}_{2}$, $\Sigma_{B}=\beta\mathbb{1}_{2}$, and $\varepsilon_{AB}=\gamma\sigma_{Z}$, we can write the average fidelity as
\begin{equation}
\overline{F}_{\text{TMST}} = \frac{1}{1+\frac{1}{2}\left(\alpha+\beta-2\gamma\right)}.
\end{equation}
If we consider the composition of $k$ teleportation protocols where each of the parties involved is separated by $L/k$, being $L$ the total distance aimed to cover. The final average fidelity is then given by
\begin{equation}
\overline{F}^{(k)}_{\text{TMST}} = \frac{1}{1+\left(k-\frac{1}{2}\right)\left(\alpha+\beta-2\gamma\right)},
\end{equation}
such that $\overline{F}_{\text{TMST}}>\overline{F}^{(k)}_{\text{TMST}}$ for $k>1$. Since the composition of teleportation protocols does not improve the overall fidelity, we study entanglement distillation and entanglement swapping in search for such gain. However, this fidelity composition may improve the overall fidelity when diffraction effects at the termination of the antenna come into play, which will reduce the reach of entanglement from the hundreds to the tens of meters.  

In Table~\ref{table1} we present the parameters we have used to represent the different fidelity curves in this section.
\begin{table}
\centering
\begin{tabular}{|l|c|c|}
\hline
Parameter &Symbol &Value\\
\hline
Losses per unit of length  &$\mu$ & $1.44\cdot 10^{-6} \text{m}^{-1}$\\
Atmospheric temperature & $T$ &  300 K\\
Mean photon number & $N_\text{th}$ & 1250\\
Squeezing parameter & $r$ &  1\\
Thermal photon number (signal) & $n$ & $10^{-2}$ \\
Transmission coefficient & $\tau$ &  0.95\\
Antenna reflectivity & $\eta_{\text{ant}}$ & 0\\
\hline
\end{tabular}
\caption{Parameters for a terrestrial (1 atm of pressure, temperature of 300 K)  two-mode squeezed thermal state generated at a 50 mK cryostat, for a frequency of 5 GHz. These parameter values correspond to an Earth-based quantum teleportation scenario.}
\label{table1}
\end{table}

\subsubsection{Asymmetric case}
Assume that Alice generates a TMST state and sends one of the modes to Bob. Then, the covariance matrix of the state, given in Eq.~\eqref{CM_asym}, is characterised by
\begin{eqnarray}
\nonumber \alpha &=& (1+2N_{\text{th}})\eta_{\text{eff}}  + (1+2n)(1-\eta_{\text{eff}})\cosh 2r, \\
\nonumber \beta &=& (1+2n)\cosh 2r, \\
\gamma &=& (1+2n)\sqrt{1-\eta_{\text{eff}}}\sinh 2r,
\end{eqnarray}
which results in an average fidelity
\begin{eqnarray}
\nonumber \overline{F}_{\text{TMST}} &=& \bigg[ 1 + \left(\frac{1}{2}+N_{\text{th}}\right)\eta_{\text{eff}}\\
&+&\left(\frac{1}{2}+n\right)(2-\eta_{\text{eff}})\cosh2r \\
&-& (1+2n)\sqrt{1-\eta_{\text{eff}}}\sinh2r \bigg]^{-1},
\end{eqnarray}
with $\eta_{\text{eff}} = 1-e^{-\mu L}(1-\eta_{\text{ant}})$. 

\subsubsection{Symmetric case}
In this case, we consider that the resource state is generated at an intermediate point between Alice and Bob, and is sent to both of them, such that now both modes are affected by the lossy and noisy channel described above. The covariance matrix of this state, presented in Eq.~\eqref{CM_sym}, is characterized by
\begin{eqnarray}
\nonumber \alpha &=& (1+2N_{\text{th}})\eta_{\text{eff}}  + (1+2n)(1-\eta_{\text{eff}})\cosh 2r, \\
\nonumber \beta &=& (1+2N_{\text{th}})\eta_{\text{eff}}  + (1+2n)(1-\eta_{\text{eff}})\cosh 2r, \\
\gamma &=& (1+2n)(1-\eta_{\text{eff}})\sinh 2r
\end{eqnarray}
where we have assumed $L_{1}=L_{2}=L/2$, and thus $\eta^{(1)}_{\text{eff}} = \eta^{(2)}_{\text{eff}} = \eta_{\text{eff}} = 1-e^{-\mu\frac{L}{2}}(1-\eta_{\text{ant}})$ . Then, the average fidelity can be written as
\begin{eqnarray}
\nonumber \overline{F}_{\text{TMST}} &=& \bigg[ 1 + (1+2N_{\text{th}})\eta_{\text{eff}}+(1+2n)(1-\eta_{\text{eff}})\cosh2r \\
&-& (1+2n)(1-\eta_{\text{eff}})\sinh2r \bigg]^{-1}.
\end{eqnarray}
Notice that, for short distances, the fidelities associated to the asymmetric and symmetric states coincide. That is, at first order in $\mu L\ll 1$, and with $\eta_{\text{ant}}=0$, 
\begin{eqnarray}
&&\overline{F}_{\text{TMST}} \approx \\
\nonumber && \left[ 1 + (1+2N_{\text{th}})\frac{\mu L}{2}+(1+2n)\left(1-\frac{\mu L}{2}\right)e^{-2r}\right]^{-1}.
\end{eqnarray}
When considering a lossy antenna, we observe higher entanglement degradation in the symmetric state due to the fact that both modes of the state are output by an antenna, whereas only one mode of the asymmetric state goes through it. Although $\sqrt{\eta_{\text{ant}}}$ can theoretically be reduced below $10^{-9}$~\cite{GonzalezRaya2020}, this leads to a slightly lower fidelity in the case of the symmetric state. In the figures appearing in this section, however, we consider $\eta_{\text{ant}}=0$ for simplicity. 

\subsection{Fidelity with photon subtraction}

If we consider a symmetric two-photon subtraction process, in which the desired resource has lost a single photon in each mode, the average fidelity becomes
\begin{widetext}
\begin{eqnarray}
\nonumber \overline{F}_{\text{2PS}} &=& \frac{1}{4}\left[ 1+\tau\frac{-\alpha\beta + (1+\gamma)^{2} +((1-\alpha)(1-\beta)-\gamma^{2})\tau}{(1+\alpha)(1+\beta)-\gamma^{2}-(\alpha\beta-(1-\gamma)^{2})\tau}\right]^{3} \times \\
&& \left[1 + \frac{(1-\alpha\beta+\gamma^{2})^{2}-(\alpha-\beta)^{2}+4\gamma^{2}+4\gamma((1-\alpha)(1-\beta)-\gamma^{2})\tau}{(1-\alpha\beta+\gamma^{2}+((1-\alpha)(1-\beta)-\gamma^{2})\tau)^{2}-(\alpha-\beta)^{2}+4\gamma^{2}}\right].
\end{eqnarray}
\end{widetext}

\begin{figure}[t]
\includegraphics[width=0.5 \textwidth]{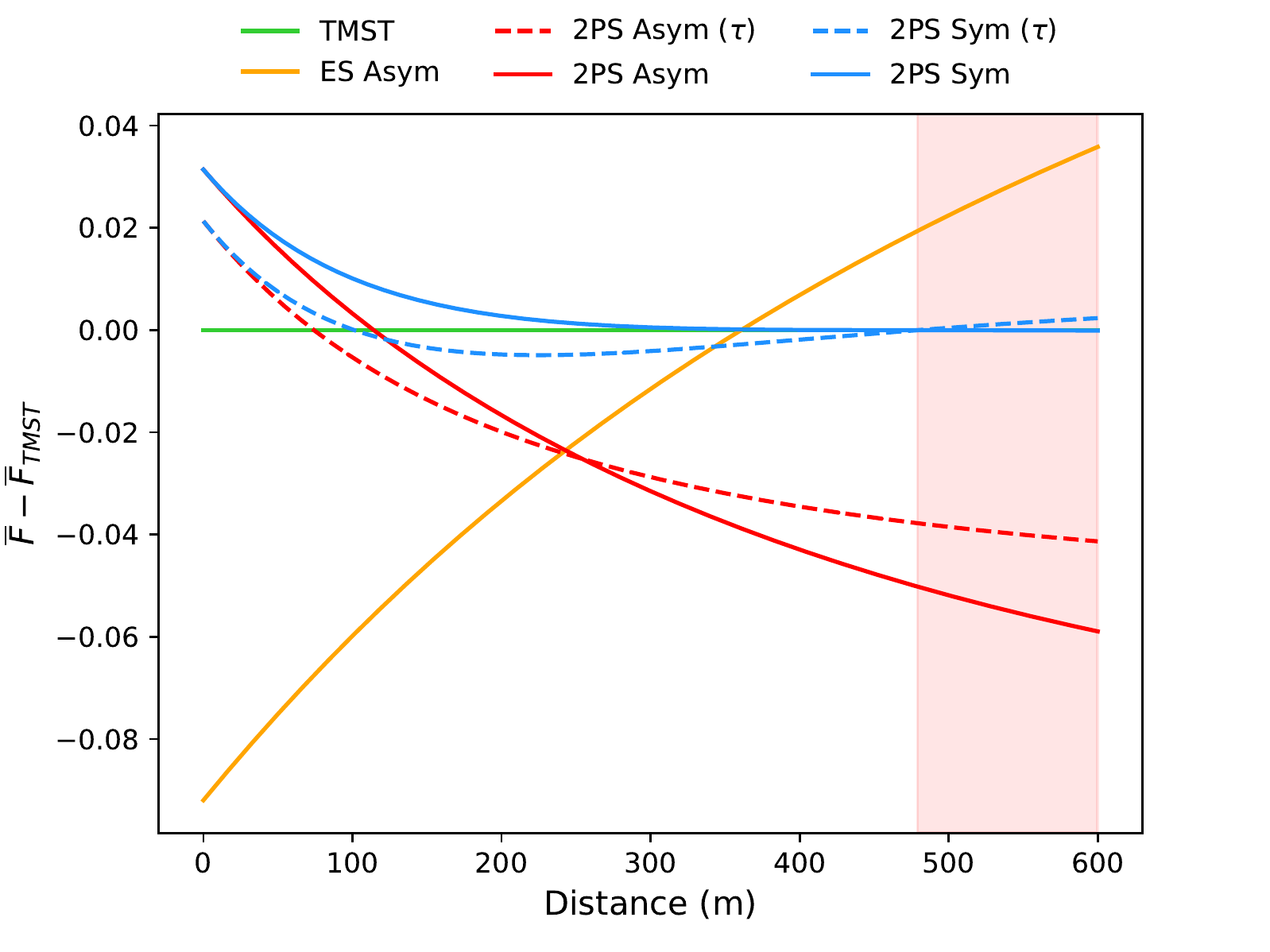}
\caption{Average fidelity of CV quantum teleportation of an unknown coherent state using an entangled resource distributed through open-air, represented versus the travelled distance. We subtract the average fidelity associated to the TMST state distributed through open-air (green), to the average fidelities of the two-photon subtracted asymmetric (2PS Asym) and symmetric (2PS Sym) states, represented in red and blue, respectively, as well as to the average fidelity of the entanglement-swapped asymmetric (ES Asym) state, in orange. We represent the states resulting from  probabilistic photon subtraction (dashed), as well as heuristic photon subtraction (solid). In a pale red background, we represent the region where the fidelity is below the maximum classical fidelity of $1/2$, and the quantum advantage is lost. The green line then shows no gain, and any point above it corresponds to an improvement in fidelity. Parameters: $n=10^{-2}$, $N_{\text{th}}=1250$, $r=1$, $\mu=1.44\cdot10^{-6}$ m$^{-1}$, $\eta_{\text{ant}}=0$, $\tau=0.95$.}
\label{fig9}
\end{figure}
\begin{figure}[t]
\includegraphics[width=0.48 \textwidth]{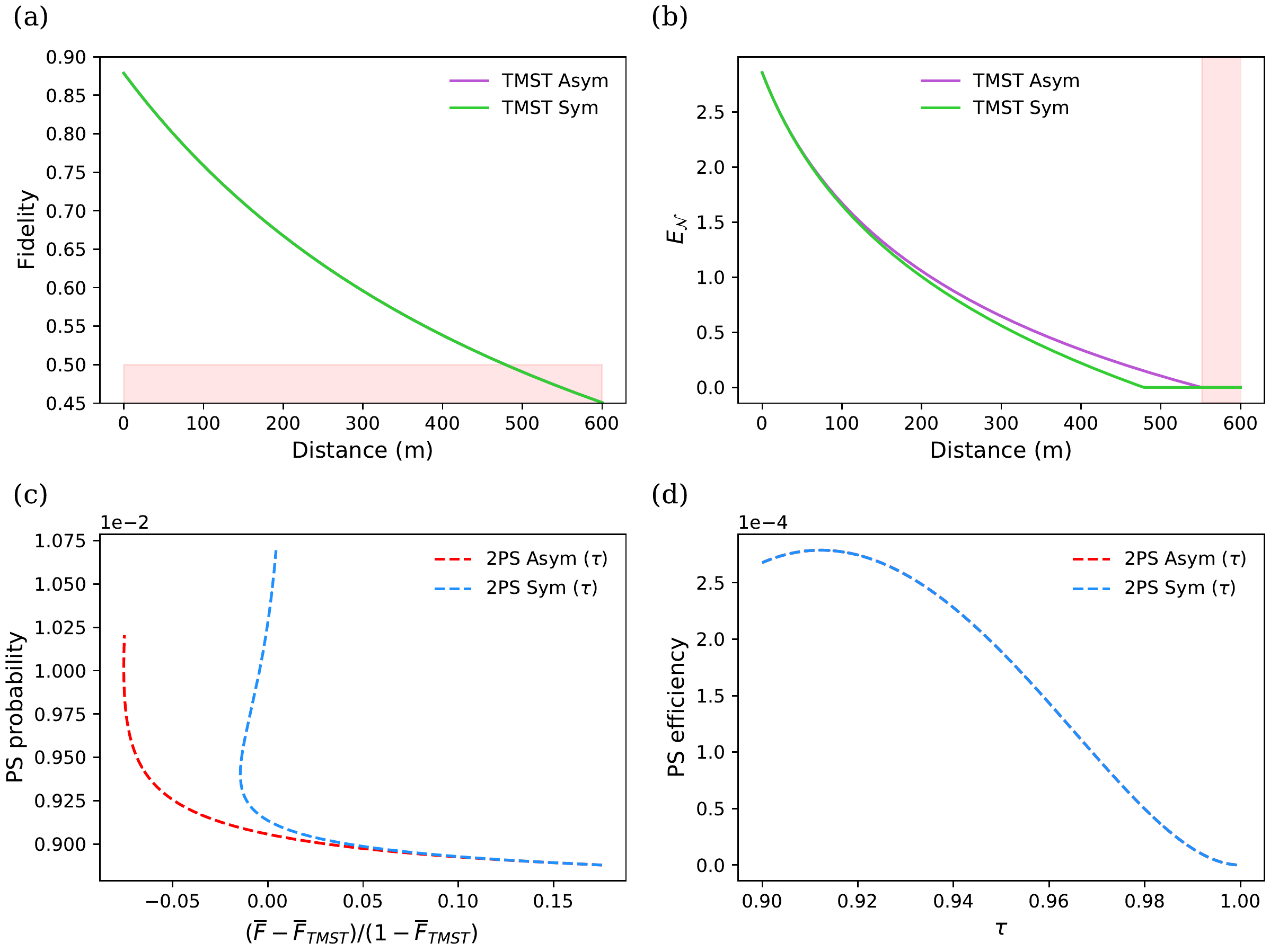}
\caption{Various features of the symmetric and asymmetric TMST states distributed through open-air. (a) Average fidelity of CV quantum teleportation protocol of an unknown coherent state using either symmetric or asymmetric lossy TMST states, represented against the travelled distance. (b) Logarithmic negativities $E_{\mathcal{N}}=\log_{2}(2\mathcal{N}+1)$ of the lossy TMST symmetric (green) and asymmetric (purple) states. (c) Probability of successful two-photon subtraction applied to lossy TMST symmetric (dashed, blue) and asymmetric (dashed, red) states, represented against the fidelity gain compared to the lossy TMST state, which is weighted to be larger for larger fidelities. (d) Efficiency of two-photon subtraction against the transmissivity, with $\tau\in[0.9,1]$, applied to TMST symmetric (dashed, blue) and asymmetric (dashed, red) states, at $x=0$. Parameters: $n=10^{-2}$, $N_{\text{th}}=1250$, $r=1$, $\mu=1.44\cdot10^{-6}$ m$^{-1}$, $\eta_{\text{ant}}=0$, $\tau=0.95$.}
\label{fig10}
\end{figure}
In Fig.~\ref{fig9} we represent the difference in fidelities associated to a CV open-air quantum teleportation protocols for an unknown coherent state, using two-mode squeezed thermal states distributed through open-air as a resource, against the travelled distance. We subtract the fidelity associated to the bare resource (TMST) to the ones related to two-photon-subtracted symmetric (blue) and asymmetric (red) states, as well as entanglement-swapped (orange) states. We consider both heuristic (solid lines) and probabilistic (dashed lines, labelled with $\tau$) photon subtraction. In Fig.~\ref{fig10}~(a), we can see the fidelity associated to the bare resource, knowing that it coincides for the symmetric and the aymmetric states in the region $\mu L\ll1$. The solid green line represents the no-gain line, above which any point represents an improvement in fidelity over the bare state. The former gives an enhancement for short distances, whereas the latter helps extend the point where the classical limit is reached. One of the reasons the gain related to photon subtraction is lost might be the increase of thermal photons in the state, which occurs for increasing $L$. This happens because, as photon losses are more relevant, the cost of doing photon subtraction is higher: if we subtract thermal photons, the entanglement hardly increases, whereas if we subtract photons from the signal, entanglement decreases. In a pale red background, we represent the region in which the fidelity associated to the bare resource reaches the maximum classical value of $1/2$.

In Fig.~\ref{fig10}, we represent various features of the two-mode squeezed thermal states distributed through open-air: (a) Average fidelity, which coincides for the symmetric and asymmetric states for $\mu L\ll 1$; (b) Logarithmic negativity $E_{\mathcal{N}}=\log_{2}(2\mathcal{N}+1)$ of the symmetric (green) and asymmetric (purple) states; (c) Sucess probability of photon subtraction (see Eq.~\eqref{eq:success-prob}) for symmetric (blue, dashed) and for asymmetric (red, dashed) states, against $(\overline{F}_{\text{2PS}}-\overline{F}_{\text{TMST}})/(1-\overline{F}_{\text{TMST}})$. This represents the gain in fidelity of the photon-subtraction schemes, weighted to show larger values when the gain occurs at larger fidelities; (d) Efficiency of photon subtraction at $x=0$, computed as $P(\overline{F}_{\text{2PS}}-\overline{F}_{\text{TMST}})$, against different values of the transmissivity, with $\tau\in[0.9,1]$. Notice that greater fidelity gains come at lower success probabilities for photon subtraction, which can be reflected in the efficiency (on the order of $10^{-4}$). The latter achieves maximum values for a transmissivity of $\tau\approx 0.92$, and goes to zero with the probability, as $\tau$ goes to 1.

In an attempt to explain the crossing that occurs between the probabilistically-photon-subtracted and the bare fidelities, which delimits the region in which photon subtraction results in an enhanced teleportation fidelity, we consider the following approach: we will attempt to find the Gaussian state that is related to our non-Gaussian photon-subtracted state by the same teleportation fidelity. Essentially, we are looking to identify the photon subtracted states with Gaussian resources in order to compute the negativities from their covariance matrices, and investigate what happens to entanglement at the points where fidelity with photon-subtracted states loses its advantage. A similar approach can be found in Appendix~\ref{app_B} for heuristic photon subtraction. First, know that the fidelity with two-photon subtraction can be written as
\begin{equation}\label{eq:fidelity_2PS}
\overline{F}_{\text{2PS}} = \frac{1+ g}{\sqrt{\det\left[\mathbb{1}_{2}+\frac{1}{2}\tilde{\Gamma}\right]}},
\end{equation}
where $\tilde{\Gamma} \equiv (\sigma_Z \tilde{\Sigma}_{A}\sigma_Z + \tilde{\Sigma}_{B} -\sigma_Z \tilde{\varepsilon}_{AB}-\tilde{\varepsilon}_{AB}^\intercal\sigma_Z)$, and $\tilde{\Sigma}_{A}$, $\tilde{\Sigma}_{B}$, and $\tilde{\varepsilon}_{AB}$ are defined in Eq.~\eqref{2PS_submatrices}. Here, $g$ is the result of integrating all the non-Gaussian corrections to the characteristic function, which enforces the non-Gaussianity of the state resulting from photon subtraction (see Appendix~\ref{app_C} for the general expression). We split the terms on the previous equation and write
\begin{widetext}
\begin{eqnarray}
\nonumber \frac{1}{\sqrt{\det\left[\mathbb{1}_{2}+\frac{1}{2}\tilde{\Gamma}\right]}} &=& \frac{1}{2}\left[ 1+\tau\frac{-\alpha\beta + (1+\gamma)^{2} +((1-\alpha)(1-\beta)-\gamma^{2})\tau}{(1+\alpha)(1+\beta)-\gamma^{2}-(\alpha\beta-(1-\gamma)^{2})\tau}\right], \\
1+g &=& \frac{1}{2}\left[ 1+\tau\frac{-\alpha\beta + (1+\gamma)^{2} +((1-\alpha)(1-\beta)-\gamma^{2})\tau}{(1+\alpha)(1+\beta)-\gamma^{2}-(\alpha\beta-(1-\gamma)^{2})\tau}\right]^{2} \times \\
&& \left[1 + \frac{(1-\alpha\beta+\gamma^{2})^{2}-(\alpha-\beta)^{2}+4\gamma^{2}+4\gamma((1-\alpha)(1-\beta)-\gamma^{2})\tau}{(1-\alpha\beta+\gamma^{2}+((1-\alpha)(1-\beta)-\gamma^{2})\tau)^{2}-(\alpha-\beta)^{2}+4\gamma^{2}}\right].
\end{eqnarray}
\end{widetext}
If we define a matrix $G = (1+g)\mathbb{1}_{2}$ with $G^{-1} = \frac{1}{1+g}\mathbb{1}_{2}$, then we can write $1+g = \sqrt{\det G}$, which leads us to 
\begin{equation}
\frac{1+ g}{\sqrt{\det\left[\mathbb{1}_{2}+\frac{1}{2}\tilde{\Gamma}\right]}} = \frac{1}{\sqrt{\det\left[\left(\mathbb{1}_{2}+\frac{1}{2}\tilde{\Gamma}\right)G^{-1}\right]}}.
\end{equation}
By rearranging the terms resulting from the matrix product, we can obtain
\begin{equation}
\left(\mathbb{1}_{2}+\frac{1}{2}\tilde{\Gamma}\right)G^{-1} = \mathbb{1}_{2} + \frac{1}{2}\left(\frac{\tilde{\Gamma} - 2g\mathbb{1}_{2}}{1+g}\right) \equiv \mathbb{1}_{2} + \frac{1}{2}\tilde{\tilde{\Gamma}},
\end{equation}
where we have defined $\tilde{\tilde{\Gamma}} = \frac{\tilde{\Gamma} - 2g\mathbb{1}_{2}}{1+g}$. Now, we want to incorporate the non-Gaussian corrections into the covariance matrix of the effective Gaussian state by using the formula
\begin{equation}
\tilde{\tilde{\Gamma}} = \sigma_{Z}\tilde{\tilde{\Sigma}}_{A}\sigma_{Z} + \tilde{\tilde{\Sigma}}_{B} - \sigma_{Z}\tilde{\tilde{\varepsilon}}_{AB} - \tilde{\tilde{\varepsilon}}_{AB}^{\intercal}\sigma_{Z}.
\end{equation}
We will refer to the resulting state as ``re-Gaussified'' state. Then, we define
\begin{eqnarray}\label{2PS_sym_eff_submatrices}
\nonumber \tilde{\tilde{\Sigma}}_{A} &=& \frac{1}{1+g}\left(\tilde{\Sigma}_{A} - g\mathbb{1}_{2}\right), \\
\nonumber \tilde{\tilde{\Sigma}}_{B} &=& \frac{1}{1+g}\left(\tilde{\Sigma}_{B} - g\mathbb{1}_{2}\right), \\
\tilde{\tilde{\varepsilon}}_{AB} &=& \frac{1}{1+g}\tilde{\varepsilon}_{AB}.
\end{eqnarray}
These represent the submatrices of a covariance matrix $\tilde{\tilde{\Sigma}}$ if
\begin{equation}
\left|\sqrt{\det\tilde{\Sigma}} - g(2+\tilde{\alpha}+\tilde{\beta}) -1\right| \geq (1+g)\left| \tilde{\alpha} - \tilde{\beta}\right|
\end{equation}
is satisfied. This condition both ensures the positivity of the covariance matrix and that the uncertainty relation is satisfied. For this, we have assumed that $\tilde{\Sigma}_{A}=\tilde{\alpha}\mathbb{1}_{2}$, $\tilde{\Sigma}_{B}=\tilde{\beta}\mathbb{1}_{2}$, and $\tilde{\varepsilon}_{AB} = \tilde{\gamma}\sigma_{Z}$. The problem is that this convention only works for the symmetric state, and not for the asymmetric one. For the latter, we write
\begin{eqnarray}
\nonumber \tilde{\tilde{\Sigma}}_{A} &=& \frac{1}{1+g}\left(\tilde{\Sigma}_{A} - kg\mathbb{1}_{2}\right), \\
\nonumber \tilde{\tilde{\Sigma}}_{B} &=& \frac{1}{1+g}\left(\tilde{\Sigma}_{B} - (2-k)g\mathbb{1}_{2}\right), \\
\tilde{\tilde{\varepsilon}}_{AB} &=& \frac{1}{1+g}\tilde{\varepsilon}_{AB}.
\end{eqnarray}
Since we have seen that a symmetric re-Gaussified state is viable, we impose the same balanced partition on the re-Gaussification of the asymmetric state. From $\tilde{\tilde{\Sigma}}_{A}=\tilde{\tilde{\Sigma}}_{B}$, we obtain that $k = 1+(\tilde{\alpha}-\tilde{\beta})/2g$, which leads to the submatrices
\begin{eqnarray}\label{2PS_asym_eff_submatrices}
\nonumber \tilde{\tilde{\Sigma}}_{A} &=& \frac{1}{1+g}\left(\frac{\tilde{\Sigma}_{A}+\tilde{\Sigma}_{B}}{2} - g\mathbb{1}_{2}\right), \\
\nonumber \tilde{\tilde{\Sigma}}_{B} &=& \frac{1}{1+g}\left(\frac{\tilde{\Sigma}_{A}+\tilde{\Sigma}_{B}}{2} - g\mathbb{1}_{2}\right), \\
\tilde{\tilde{\varepsilon}}_{AB} &=& \frac{1}{1+g}\tilde{\varepsilon}_{AB}.
\end{eqnarray}
The condition these terms need to satisfy is
\begin{equation}
\left|\sqrt{\det\tilde{\Sigma}} + \frac{1}{4}(\tilde{\alpha}-\tilde{\beta})^{2} -g(\tilde{\alpha}+\tilde{\beta}) + g^{2} - 1 \right| \geq 0,
\end{equation}
which is naturally met. In Appendix~\ref{app_D}, a graphical proof that these conditions are met is provided.

\begin{figure}
\includegraphics[width=0.48 \textwidth]{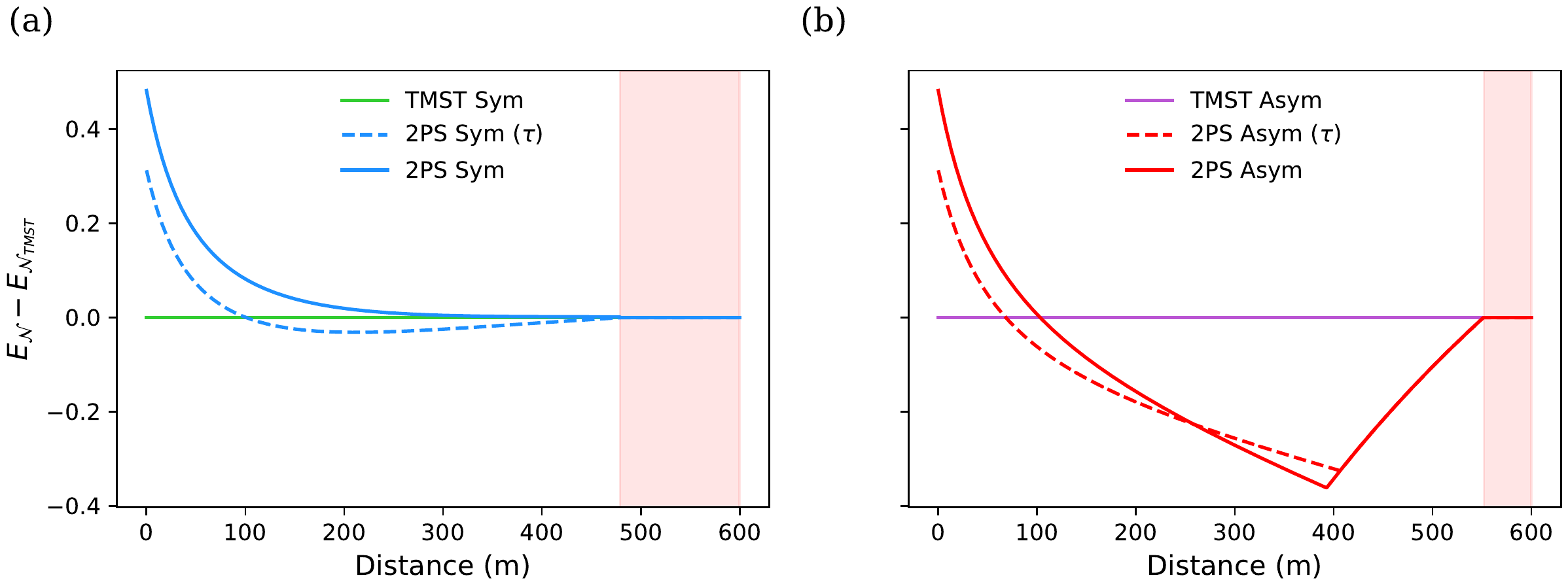}
\caption{Logarithmic negativity $E_{\mathcal{N}}=\log_{2}(2\mathcal{N}+1)$ of entangled resources distributed through open-air, represented against the travelled distance. (a) We subtract the logarithmic negativity of the lossy TMST symmetric (TMST Sym, in green) state to those of the probabilistic (dashed) and heuristic (solid) re-Gaussified two-photon subtracted symmetric states (2PS Sym, in blue). (b) We subtract the logarithmic negativity of the lossy TMST asymmetric (TMST Asym, in purple) state to those of the probabilistic (dashed) and heuristic (solid) re-Gaussified two-photon subtracted asymmetric states (2PS Asym, in red). Any point above the green (purple) line represents an improvement in negativity for the re-Gaussified photon-subtracted symmetric (asymmetric) states. Parameters: $n=10^{-2}$, $N_{\text{th}}=1250$, $r=1$, $\mu=1.44\cdot10^{-6}$ m$^{-1}$, $\eta_{\text{ant}}=0$, $\tau=0.95$.}
\label{fig11}
\end{figure}

By doing these redefinitions, we are effectively masking the non-Gaussian corrections in the expression of the fidelity as further corrections to the submatrices of the covariance matrix of an entangled resource, which is now Gaussian, while maintaining the same fidelity we had obtained with the photon-subtracted states. This treatment has shown that we are using a resource that, in the regions in which photon subtraction is beneficial, shows higher entanglement than the bare resource, as expected. In Fig.~\ref{fig11}, we subtract the logarithmic negativity $E_{\mathcal{N}}=\log_{2}(2\mathcal{N}+1)$ of the bare resource (TMST) to those of the heuristic (solid) and the probabilistic (dashed) two-photon subtracted states. On Fig.~\ref{fig11}~(a), we display the symmetric states, and on Fig.~\ref{fig11}~(b), the asymmetric ones. Notice that the gain in negativity is lost around the same points as the gain in fidelity. As we have discussed before, the fidelities corresponding to the symmetric and asymmetric states are equal at first order in $\mu L\ll 1$, and the same behavior can be observed initially in the negativities of both states (see Fig.~\ref{fig10}~(b)). However, while the points at which the fidelities of the symmetric and asymmetric states reach the classical limit differs by centimeters, the points at which entanglement is lost for these states differ by tens of meters. This region where negativity is lost is shown in a pale red background. Although the entanglement in the asymmetric state reaches further, the symmetric photon subtraction protocol we envision works better when applied on the symmetric state. The logarithmic negativity of heuristic photon-subtracted states presents a 46\% increase with respect to the value for the bare state at $x=0$, while probabilistic photon-subtracted states only present an initial gain of 28\%.

\subsection{Fidelity with entanglement swapping}
We consider the case in which both Alice and Bob produce two-mode squeezed states, and each send one mode to Charlie, who is equidistantly-located from the two parties. Then, he performs entanglement swapping using the two modes he has received, which have been degraded by thermal noise and photon losses. If Alice and Bob use the remaining entangled resource they share for teleporting an unknown coherent state, the fidelity of the protocol will be given by
\begin{equation}
\overline{F}_{\text{es}} = \frac{1}{1+\alpha-\frac{\gamma^{2}}{\beta}}, 
\end{equation}
where now we have
\begin{eqnarray}
\nonumber \alpha &=& (1+2n)\cosh 2r, \\
\nonumber \beta &=& (1+2N_{\text{th}})\eta_{\text{eff}}  + (1+2n)(1-\eta_{\text{eff}})\cosh 2r, \\
\gamma &=& (1+2n)\sqrt{1-\eta_{\text{eff}}}\sinh 2r,
\end{eqnarray}
and $\eta_{\text{eff}} = 1-e^{-\mu L/2}(1-\eta_{\text{ant}})$, since the total distance has been reduced in half by the presence of a third, equidistant party. 

This fidelity is represented as the orange curve in Fig.~\ref{fig9}, where it shows a gain in fidelity for large distances, right before the classical limit of $\overline{F}=0.5$ is reached. The extended distance represents 14\% of the maximum distance for the bare TMST state. This will be advantageous when the distance at which the classical limit occurs can be extended, for example in the case of quantum communication between satellites.  

\section{Inter-Satellite Quantum Communication Model}\label{section:satellites}
An important application of the protocols and the technology addressed in this manuscript would be quantum communication between satellites \cite{Sidhu2021, Pirandola2021, Pirandola2021_2}. Given the security inherent to quantum-based communication protocols, many of the motivations for the use of sub-millimiter microwaves --i.e. frequencies in the range 30-300 GHz, which is a trend in classical communication between satellites orbiting low Earth orbits (LEOs), fade away, and it seems reasonable to aim at maximizing the distances between linked satellites~\cite{Sanz2018}.  
We shall consider a greatly simplified model for free-space microwave communication, assuming unpolarized signals and hence ignoring the effects of scintillation and polarization rotation, among others. This means that whenever we discuss entanglement, it will be understood that we are talking about particle number entanglement. Polarization entanglement, even if perhaps more natural when considering the physics of antennae, is lost whenever the signal enters a coplanar waveguide, hence making it not a good candidate for quantum communication between 1D-superconducting chips. Moreover, we will assume that the communication is done within the same altitude, i.e. that the two satellites are in similar orbits, which is typically the case when building satellite constellations. This means that the atmospheric absorption, if any, will remain constant during the time of fly of the signals. Additionally, we will ignore Doppler effects caused by relative speeds between the orbits.

There are four main families of satellite orbits: GEO, HEO, MEO, and LEO, corresponding to geosynchronous, high, medium, and low Earth orbits, respectively. It is customary to define LEOs as orbits with altitudes in the range 700-2000 km; MEOs would then range between 2000-35786 km; and HEOs in 35786-$d_M/2$, where $d_M$ is the distance from the Earth to the Moon. The seemingly arbitrary altitude separating MEOs and HEOs is actually the average altitude for which the period equals one sidereal day (23h 56m 4s), and this is precisely where GEOs sit. This altitude is more than three times the point at which the exosphere, the last layer of the atmosphere, is observed to fade. GEOs and HEOs are hence `true' free-space orbits, in the sense that there is hardly any gas, and temperature is dominated by the Cosmic Microwave Background --that peaks at 2.7 K. The MEO region is the least populated one, since it is home for the Van Allen belts, which contain charged particles moving at relativistic speeds due to the magnetic field of the Earth, and that can destroy unshielded objects. LEOs, on the other hand, are `cheap' orbits, where most of the satellites orbiting our planet live. Their low altitudes simplify the problems arising from delays between Earth-based stations and the satellites.

In this section we will be concerned only with two satellites orbiting either the same GEO or the same LEO, as a simple case study of expected losses and entanglement degradation. There are essentially two kinds of loss one must take into account:  atmospheric and free-space path loss (FSPL). Total loss will be then simply given by
\begin{equation}
    L = L_\text{A} L_{\text{FSPL}}.
\end{equation}

Atmospheric absorption loss is caused by light-matter interactions. These strongly depend on the altitude of the orbits considered, among other parameters such as polarization, frequency, or wheather conditions. Atmospheric loss can range from almost negligible (up in the exosphere and beyond), to very significant in the lower layers of the atmosphere, especially when water droplets and dust are present. Atmospheric loss has to be taken into account when considering the case of up and downlinks, i.e. when linking a satellite with an Earth-based station. However, for relatively high altitudes --that is, any altitude where there are satellites, absorption loss is so low in microwaves that it can be taken to vanish as a first approximation, so we set $L_\text{A} =1$.

FSPL is due to the inevitable spreading of a signal in three dimensions; they are often referred to as geometric losses. FSPL is maximal when there is no beam-constraining mechanism, such as a wave guide, or a set of focalising lenses, i.e. when the signal spreads isotropically: $L_\text{FSPL}=\left({\lambda}/{4\pi d } \right)^{-2}$.

Suppose two co-moving satellites are separated by a linear distance $d$, and the emitter sends a quasi-monochromatic signal with power $P_e$ centered at frequency $\nu=\omega/2\pi = c/\lambda$. The receiver gets a power $P_r$ such that their ratio defines a transmission coefficient that is the product of loss and the gains (or directivities) of the antennae. The resulting equation for long, `far-field' distances is sometimes referred to as Friis' equation \cite{Friis1946, Hogg1993, Shaw2013}, which is the compromise between gain (or directivity) and loss:
\begin{equation}
\frac{P_e}{P_r} =\frac{D_e D_d}{L_\text{A} L_{\text{FSPL}}}= D_e D_r  \left(\frac{\lambda}{4\pi d } \right)^2 \equiv\breve{\tau}_\text{path},
\end{equation}
where $D_e$ and $D_r$ are the directivities of the emitter and receiver antennas, and we set $L_\text{A}=1$ as discussed before. The directivity of an antenna is the maximized gain in power in some preferred direction with respect to a hypothetical isotropic antenna, at a fixed distance from the source, and assuming the total radiation power is the same for both antennas: $D = \max_{\theta, \phi} D(\theta, \phi)$. It is a quantity that strongly depends on the geometry design, but that can be enhanced in a discrete fashion by means of antenna arrays. Indeed, given $N$ identical antennas with directivity gain $D(\theta, \phi)$, a phased array consists of an array of such antennas, each preceded by a controlled phase shifter. This diffraction problem essentially gives $D_\text{array}(\theta, \phi) = A^2_N(\bm{\varepsilon}) D(\theta, \phi)$, where $A_N$ is the so-called $N$-array factor, that symbolically depends on the phases via some vector $\bm{\varepsilon}$~\cite{Balanis2005}. In three dimensions, phase arrays are two-dimensional grids of antennas, so that the main lobe of the resulting signal becomes as sharp as possible. We will assume that we have an array of small coplanar antennas as the one discussed in Section~\ref{subsection:antenna_model}, adding up to a radiation pattern mimicking that of a parabolic antenna. We also assume that both emitter and receiver have the same design, $D_e = D_r \equiv D$:
\begin{equation}
D=\left(\frac{\pi a}{\lambda}\right)^2 e_a
\end{equation}
where $0 \leq e_a \leq 1$ is the aperture efficiency, defined as the ratio between the effective aperture $A_e$, and the area of the antenna's actual aperture, $A_\text{phys}$, and $a$ is the diameter of the parabola, such that $A_{\text{phys}}=\pi a^{2}/4$. With this, the parabolic path transmissivity becomes
\begin{equation}
\breve{\tau}_\text{path} = \left( \frac{\pi a^2 e_a }{4 d \lambda}\right)^{2}.
\end{equation}
\begin{figure}
\includegraphics[width=0.48 \textwidth]{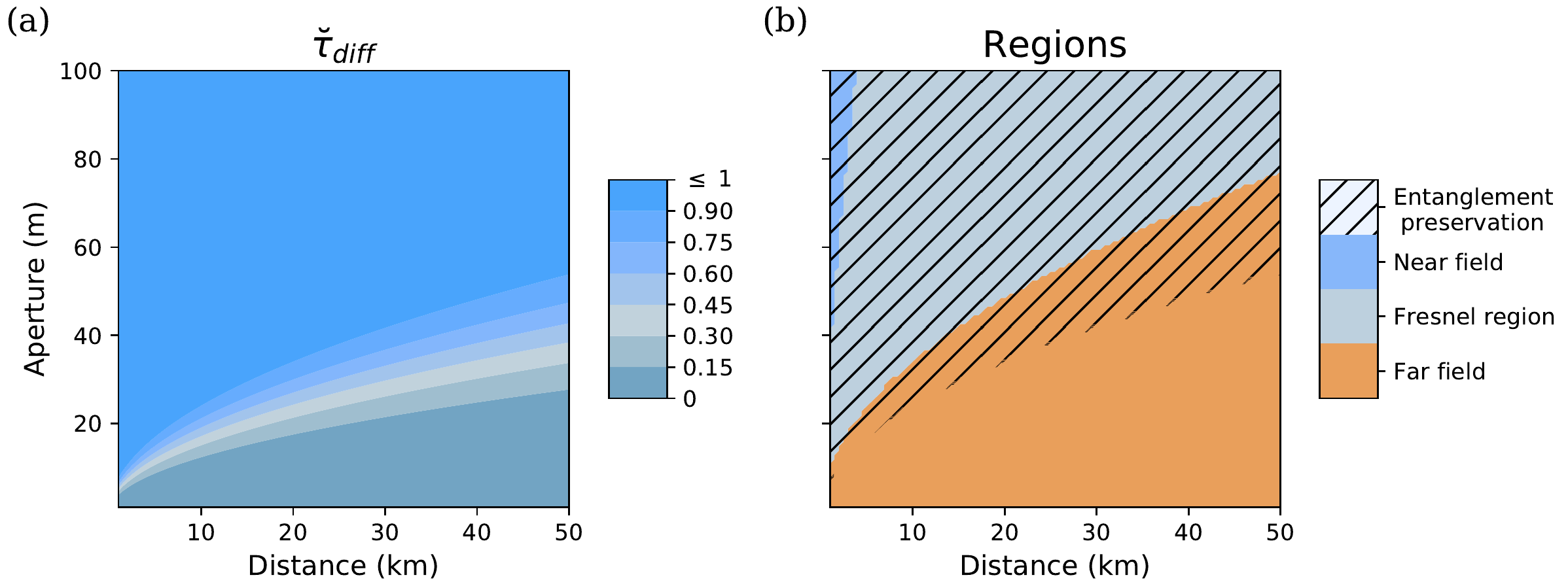}
\caption{(a) Contour plot of the transmissivity associated to diffraction, $\breve{\tau}_\text{diff}$, against the aperture radius of the antenna and the travelled distance. We can observe that losses are greatly reduced with the aperture of the antenna. (b) Contour plot of the regions of free-space as delimited by the relation between the aperture radius of the antenna and the distance at which the signal is observed: near-field (blue) $\varpi_{0} > (\lambda d/0.62)^{2/3}$, Fresnel (grey) $\sqrt{\lambda d/2} < \varpi_{0} < (\lambda d/0.62)^{2/3}$, and far-field (orange) $\varpi_{0} < \sqrt{\lambda d/2}$. With dashed lines, we represent the region where entanglement can be preserved. Parameters: $\lambda = 6 \text{ cm}$, $a_{R}=2\varpi_{0}$.}
\label{fig12}
\end{figure}
The effect of path losses can alternatively be described by a diffraction mechanism, affecting the spot size of the signal beam,
\begin{equation}
\varpi = \frac{\varpi_{0}}{\sqrt{2}}\sqrt{\left(1-\frac{d}{R_{0}}\right)^{2} + \left(\frac{d}{d_{R}} \right)^{2}},
\end{equation}
given an initial spot size $\varpi_{0}$, curvature of the beam $R_{0}$, and Rayleigh range $d_{R}=\pi\varpi_{0}^{2}\lambda^{-1}/2$. Given the aperture radius $a_{R}$ of the receiver antenna, the diffraction-induced transmissivity can be computed as~\cite{Pirandola2021,Pirandola2021_2}
\begin{equation}
\breve{\tau}_{\text{diff}} = 1 - e^{-2a_{R}^{2}/\varpi^{2}}.
\end{equation}
Notice that, in the far-field approximation, we can recover the result for $\breve{\tau}_\text{path}$, 
\begin{equation}
\breve{\tau}_{\text{diff}} \approx \left( \frac{\pi\varpi_{0}a_{R}}{\lambda d}\right)^{2},
\end{equation}
by setting $a_{R}=\varpi_{0}=a/2$, $R_{0}=d$, and assuming $e_{a}=1$. Setting $\lambda = 6 \text{ cm}$ and $a_{R}=2\varpi_{0}$, we plot the transmissivity associated to diffraction vs the distance $d$ for different values of the aperture $\varpi_{0}$ in Fig.~\ref{fig12}~(a), observing that losses are reduced as a result of an increase of the aperture.

We address entanglement preservation in TMST states distributed through open-air by considering that the dominant source of error will be diffraction, as opposed to attenuation, which we will describe by means of a beamsplitter with a thermal input. We introduce $N_{\text{th}}\sim 11$ as the number of thermal photons in the environment at $2.7$ K. Considering this loss mechanism, entanglement preservation is achieved for reflectivities that satisfy $\eta<(1+N_{\text{th}})^{-1}\sim0.083$ for lossy TMST asymmetric states, and $\eta<[1+N_{\text{th}}(1+\coth r)]^{-1}\sim0.038$ for symmetric ones, assumming $n\approx 0$ and $\breve{\tau}=1-\eta$. Given this diffraction channel, entanglement is preserved in the regime $a_{R}\varpi_{0}/d > (\lambda/\pi)\sqrt{-\log\eta_{\text{lim}}} \sim 0.035$, for $\lambda=6$ cm and $\eta_{\text{lim}}=0.038$. This implies that, for two satellites that are are separated by $d=1$ km, the product of apertures of emitter and receiver antennae must be $a_{R}\varpi_{0} > 35 \text{ m}^{2}$ in order to have entanglement preservation. In Fig.~\ref{fig12}~(b), we represent the regions of free-space as delimited by the relation between the distance at which the signal is detected and the aperture of the emitting antenna, taking $a_{R}=2\varpi_{0}$, and depicting the region in which entanglement is preserved with a dashed line. This shows that the radius of the antennae of emmitter and receiver satellites will be large, as is usually the case for microwave communications. In order to correct the effects of diffraction with microwaves, it would also be useful to study focalizing techniques and the incorporation of beam collimators.

\section{Conclusions}
In this manuscript, we have studied the feasibility of microwave entanglement distribution in open air with two-mode squeezed states. We have studied these as a resource for a Braunstein-Kimble quantum teleportation protocol adapted to microwave technology, reviewing the steps involved in this process and the possibility of realization, given the available quantum technology. First, we reviewed the process of generating two-mode squeezed states using JPAs, which was tested in Ref.~\cite{Fedorov2018}. Then, we looked at an antenna model~\cite{GonzalezRaya2020} for optimal transmission of these states into open-air, while also discussing entanglement degradation due to interaction with the environment and the maximum possible reach of entanglement. This was found to be up to 550 meters for asymmetric states with ideal weather conditions. Following this, we have have adapted to microwave technology entanglement distillation and entanglement swapping, two procedures to counteract degradation at different stages. In particular, we studied photon subtraction, an entanglement distillation protocol that works for short distances and low squeezing, allowing for up to 46\% increase in the logarithmic negativity. Entanglement distillation, on the other hand, contributed to extending the reach of entanglement by up to 14\%. Since these operations will require homodyne detection, as well as photocounting, we discussed recent advances in microwave technologies that permit these operations. We then tested the efficiency of open-air entanglement distribution, including the different enhancement techniques, with the Braunstein-Kimble protocol adapted to microwaves: we computed the average fidelities of open-air microwave quantum teleportation of coherent states using the open-air distributed states as the entangled resource. We concluded with a study of the applicability and efficiency of these techniques for quantum communication between satellites, a field where, with the proper directivity, given the low absorption rates, the reach of entanglement can be greatly increased. 

Efficient information retrieval from open-air distribution of microwave quantum states will be a key step in this protocol, which would require the design of a receiver antenna. This device may resemble the one described in Section~\ref{subsection:antenna_model}, but it will have to include a different termination into open-air in order to, for instance, reduce diffraction losses. Since the lack of an amplification protocol considerably limits the entanglement transmission distances through open-air, a hopeful solution is to develop a theory of quantum repeaters for microwave signals, following the ideas shown in Ref.~\cite{DiCandia2015}. For this, entanglement distillation and entanglement swapping techniques discussed in this manuscript will be useful. 

Since superconducting circuits naturally work in the microwave regime, it would make sense to explore alternatives for photon subtraction that use devices specific to this technology. In that direction, a possible deterministic photon-subtraction scheme could be studied, making use of circuit-QED for non-demolition detection of itinerant microwave photons~\cite{Lescanne2020}. In this work, the detection of a previously unknown microwave photon is guaranteed by a transmon qubit jumping to its excited state, which would indicate a successful photon subtraction event.

\acknowledgements
The authors thank R. Assouly, R. Dassonneville and B. Huard for useful discussions. 
\\
All authors acknowledge support from QMiCS (Grant No. 820505) of the EU Flagship on Quantum Technologies. 
TG-R and MS acknowledge financial support from the QUANTEK project from ELKARTEK program (KK-2021/00070), as well as from OpenSuperQ (820363) of the EU Flagship on Quantum Technologies, and the EU FET-Open projects Quromorphic (828826) and EPIQUS (899368).
MC and YO thank the support from Funda\c{c}\~{a}o para a Ci\^{e}ncia e a Tecnologia (Portugal), namely through project UIDB/50008/2020, as well as from project TheBlinQC supported by the EU H2020 QuantERA ERA-NET Cofund in Quantum Technologies and by FCT (QuantERA/0001/2017). MC acknowledges support from the DP-PMI and FCT through scholarship PD/BD/135186/2017. 
MR, FF, FD, and KF acknowledge support by the German Research Foundation via Germany’s Excellence Strategy (EXC-2111-390814868), Elite Network of Bavaria through the program ExQM, and the German Federal Ministry of Education and Research via the project QUARATE (Grant No. 13N15380) and the project QuaMToMe (Grant No. 16KISQ036). This research is part of the Munich Quantum Valley, which is supported by the Bavarian state government with funds from the Hightech Agenda Bayern Plus.
The research of VS is supported by the Basque Government through the BERC 2022-2025 program and by the Ministry of Science, Innovation, and Universities: BCAM Severo Ochoa accreditation SEV-2017-0718.

MM acknowledges funding from the European Research Council under Consolidator Grant No. 681311 (QUESS), and the Academy of Finland through its Centers of Excellence Program(project Nos. 312300, and 336810).



\appendix

\section{Amplification}\label{app_A}
In this manuscript, we have not discussed an amplification protocol for entangled signals because it cannot increase quantum correlations. Amplification of signals is an essential feature in classical microwave communication in open-air, as increasing the number of photons that Alice sends will improve the chances Bob has of detecting that signal. In the quantum regime, cryogenic high electronic mobility transistor (HEMT) amplifiers are suited for experiments with microwaves. These are able to greatly enhance signals in a large frequency spectrum, while introducing a significant amount of thermal photons. This noise is reflected in the input-output relation 
\begin{equation}
a_{\text{out}}=\sqrt{g_H}a_{\text{in}}+\sqrt{g_H-1}h_H,
\end{equation}
with $a_{\text{in}}$, $a_{\text{out}}$, and $h_H$ the annihilation operators of the input field, output field and noise added by the amplifier, respectively. From this formula, we can see that amplification is a procedure that acts individually on the modes of a quantum state, which means that we can increment the number of photons of that mode, but they will not be entangled with the other ones. That is why we say that amplification cannot increase quantum correlations. If anything, the introduction of thermal noise can lead to entanglement degradation. 

HEMTs normally work at 4K temperatures, which implies that the number of thermal photons they introduce is around $n_{H} \sim 10 - 20$, for 5 GHz frequencies. It is the number of thermal photons which determines the thermally-radiated power~\cite{Clerk2010}, $P = \hbar \omega N B$, meaning that the excess output noise produces a flux of $N$ photons per second in a bandwidth of $B$ Hz. The gain is given by the ratio between output and input powers, and for a constant bandwidth, it is just $g_{H}=n_{H}/n$, the ratio between the number of thermal photons introduced by the HEMT and the number of photons in the input state. Considering the antenna described above, an HEMT of these characteristics produces a gain of $g_{H}=2\cdot 10^{3}$ when acting on a TMST state with $n\sim 10^{-2}$, which completely destroys entanglement. In order for entanglement to survive such an amplification process, the HEMT must be placed at temperatures below 100 mK. In any case, they do not present an advantage. 

\section{Heuristic photon subtraction}\label{app_B}
In this section, we describe the heuristic photon subtraction operation, in which annihilation operators are applied to a quantum state in order to reduce the number of photons in each mode. Consider that we have initially a two-mode squeezed vacuum (TMSV) state,
\begin{equation}
|\psi\rangle_{AB} = \sqrt{1-\lambda^{2}}\sum_{n=0}^{\infty}\lambda^{n}|n,n\rangle_{AB},
\end{equation}
with $\lambda=\tanh r$ and squeezing parameter $r$. Assume that we apply the operator $a_{A}^{k}a_{B}^{l}$, which implies subtracting $k$ photons in mode $A$ and $l$ photons on mode $B$. The resulting state is 
\begin{eqnarray}
&& |\psi^{(k+l)}\rangle_{AB} = \sqrt{1-\lambda^{2}}\times \\
\nonumber &&\sum_{n=\max\{k,l\}}^{\infty}\frac{\lambda^{n}n!}{\sqrt{(n-k)!(n-l)!}}|n-k,n-l\rangle_{AB}.
\end{eqnarray}
Considering symmetric photon subtraction ($k=l$), we can write this state
\begin{equation}
|\psi^{(2k)}\rangle_{AB} = \sqrt{1-\lambda^{2}}\sum_{n=0}^{\infty}\frac{\lambda^{n+k}(n+k)!}{n!}|n,n\rangle_{AB},
\end{equation}
where we have shifted $n\rightarrow n+k$. This state needs to be normalized, and the normalization is given by
\begin{eqnarray}
N_{2k} &=& (1-\lambda^{2})\lambda^{2k}(k!)^{2}\sum_{n=0}^{\infty}\lambda^{2n}\begin{pmatrix}n+k \\ k\end{pmatrix}^{2} \\
\nonumber &=& (1-\lambda^{2})\lambda^{2k}(k!)^{2} \, _2F_1\left(k+1,k+1;1; \lambda^2\right).
\end{eqnarray}
The negativity of this state can be computed as
\begin{equation}
\mathcal{N}(\rho^{(2k)}) = \frac{A_{k}-1}{2},
\end{equation}
with $\rho^{(2k)}=|\psi^{(2k)}\rangle\langle\psi^{(2k)}|$ and
\begin{equation}
A_{k} \equiv \left( \sqrt{\frac{1-\lambda^{2}}{N_{2k}}}\sum_{n=0}^{\infty} \lambda^{n+k}\frac{(n+k)!}{n!}\right)^{2}.
\end{equation}
This leads to
\begin{equation}
\mathcal{N}(\rho^{(2k)}) = \frac{1}{2} \left(\frac{(1-\lambda)^{-2 (k+1)}}{\, _2F_1\left(k+1,k+1;1; \lambda^2\right)}-1\right)
\end{equation}
from which we can recover the negativity of the TMSV state by setting $k=0$,
\begin{equation}
\mathcal{N}(\rho^{(0)}) = \frac{\lambda}{1-\lambda} = \frac{e^{2r}-1}{2}.
\end{equation}

Now, we go beyond the TMSV state case, and explore the heuristic photon-subtraction protocol applied on a general Gaussian state. The application of the single-photon annihilation operators on both modes of a bipartite quantum state $\rho$ is equivalent to applying
\begin{equation}
\Theta_{1}\Theta_{2}\chi(\alpha,\beta)
\end{equation}
to its the characteristic function, with
\begin{equation}
\Theta_{i} = \partial_{x_{i}^{2}} + \partial_{p_{i}^{2}} + \frac{x_{i}^{2}}{4} + \frac{p_{i}^{2}}{4} + x_{i}\partial_{x_{i}} + p_{i}\partial_{p_{i}} + 1,
\end{equation}
for $i=\{1,2\}$. Given that $\rho = \frac{1}{\pi^{2}}\int \diff^{2}\alpha\int \diff^{2}\beta \chi(\alpha,\beta)D_{1}(-\alpha)D_{2}(-\beta)$, and assuming that $\rho$ is a Gaussian state with covariance matrix 
\begin{equation}
\Sigma = \begin{pmatrix} \Sigma_{A} & \varepsilon_{AB} \\ \varepsilon^{\intercal}_{AB} & \Sigma_{B}\end{pmatrix},
\end{equation}
then we can write
\begin{widetext}
\begin{eqnarray}
&& \Theta_{1}\Theta_{2}\chi(\alpha,\beta) = N \Big[ \left( m_{B} + \vec{\beta} M_{B} \vec{\beta}^{\intercal} + \vec{\alpha}M_{BC}\vec{\beta}^{\intercal} + \vec{\alpha} M_{C} \vec{\alpha}^{\intercal} \right) \left( m_{A} + \vec{\alpha} M_{A} \vec{\alpha}^{\intercal} + \vec{\alpha}_{1}M_{AC}\vec{\beta}^{\intercal} + \vec{\beta} M_{C} \vec{\beta}^{\intercal} \right) \\
\nonumber && + m_{C} - \vec{\alpha}M_{AC}\Omega \varepsilon_{AB}\Omega^{\intercal}\vec{\alpha}^{\intercal} + 2\vec{\beta}M_{C}\left(\mathbb{1}_{2} - \Omega \Sigma_{B}\Omega^{\intercal}\right)\vec{\beta}^{\intercal} + \vec{\alpha}\left[M_{AC}\left(\mathbb{1}_{2}-\Omega \Sigma_{B}\Omega^{\intercal}\right) - \Omega\varepsilon_{AB}\Omega^{\intercal}M_{C}\right]\vec{\beta}^{\intercal} \Big] \chi(\alpha,\beta),
\end{eqnarray}
\end{widetext}
where we have defined 
\begin{eqnarray}
\nonumber m_{A} &=& 1 - \frac{1}{2}\tr \Sigma_{A}, \\
\nonumber m_{B} &=& 1 - \frac{1}{2}\tr \Sigma_{B}, \\
\nonumber m_{C} &=& \frac{1}{2}\tr \varepsilon_{AB}^{2}, \\
\nonumber M_{A} &=& \frac{1}{4}\left( \mathbb{1}_{2} -2\Omega\Sigma_{A}\Omega^{\intercal} + \Omega \Sigma_{A}^{2}\Omega^{\intercal}\right), \\
\nonumber M_{B} &=& \frac{1}{4}\left( \mathbb{1}_{2} -2\Omega \Sigma_{B}\Omega^{\intercal} + \Omega \Sigma_{B}^{2}\Omega^{\intercal}\right), \\
\nonumber M_{C} &=& \frac{1}{4}\Omega \varepsilon_{AB}^{2}\Omega^{\intercal}, \\
\nonumber M_{AC} &=& \frac{1}{2}\left( \Omega \Sigma_{A}\varepsilon_{AB}\Omega^{\intercal} - \Omega\varepsilon_{AB}\Omega^{\intercal} \right), \\
M_{BC} &=& \frac{1}{2}\left( \Omega\varepsilon_{AB}\Sigma_{B}\Omega^{\intercal} - \Omega\varepsilon_{AB}\Omega^{\intercal}\right).
\end{eqnarray}
Here, $N^{-1}=m_{A}m_{B}+m_{C}$ is a normalization constant, which enforces $\Theta_{1}\Theta_{2}\chi(\alpha,\beta)|_{\alpha=0,\beta=0}=1$. Keep in mind that we have assumed that the submatrices $\Sigma_{A}$, $\Sigma_{B}$, and $\varepsilon_{AB}$ of the covariance matrix are symmetric.

If we use these states as the entangled resource on a CV quantum teleportation protocol of an unknown coherent state, we get that the average fidelity can be written as
\begin{equation}
\overline{F} = \frac{1+ h}{\sqrt{\det\left[\mathbb{1}_{2}+\frac{1}{2}\Gamma\right]}},
\end{equation}
with $\Gamma \equiv \sigma_Z \Sigma_{A}\sigma_Z + \Sigma_{B} -\sigma_Z \varepsilon_{AB}-\varepsilon_{AB}^\intercal\sigma_Z$. Here, we have defined $h$ as
\begin{widetext}
\begin{eqnarray}
\nonumber h &=& \frac{1}{E_{0}}\bigg\{\tr\left[\Omega\left(\mathbb{1}_{2}+\frac{1}{2}\Gamma\right)^{-1}\Omega^{\intercal}E_{1}\right] -\frac{2}{\det\left(\mathbb{1}_{2}+\frac{1}{2}\Gamma\right)}\tr\left(\Omega E_{2}^{A}\Omega^{\intercal}E_{2}^{B}\right) \\
&+& 3\tr\left[ \Omega\left(\mathbb{1}_{2}+\frac{1}{2}\Gamma\right)^{-1}\Omega^{\intercal}E_{2}^{A}\right]\tr\left[ \Omega\left(\mathbb{1}_{2}+\frac{1}{2}\Gamma\right)^{-1}\Omega^{\intercal}E_{2}^{B}\right]\bigg\},
\end{eqnarray}
\end{widetext}
together with
\begin{eqnarray}
\nonumber E_{0} &=& m_{A}m_{B} + m_{C}, \\
\nonumber E_{1} &=& m_{A}\left( M_{B} + \sigma_{Z}M_{C}\sigma_{Z}+ \sigma_{Z}M_{BC}\right) \\
\nonumber &+& m_{B}\left( \sigma_{Z}M_{A}\sigma_{Z} + M_{C} + \sigma_{Z}M_{AC}\right) \\
\nonumber &+& \left(2M_{C}+\sigma_{Z}M_{AC}\right)\Omega\left( \mathbb{1}_{2} + \sigma_{Z}\varepsilon_{AB} - \Sigma_{B} \right)\Omega^{\intercal} \\
\nonumber E_{2}^{A} &=& M_{C} + \sigma_{Z}M_{AC} + \sigma_{Z}M_{A}\sigma_{Z}, \\
E_{2}^{B} &=& M_{B} + \sigma_{Z}M_{BC} + \sigma_{Z}M_{C}\sigma_{Z}.
\end{eqnarray}
We can identify $h$ as the non-Gaussian corrections to the fidelity, and mask them as corrections to the covariance matrix of a Gaussian state with the same fidelity. This is done by defining 
\begin{equation}
\tilde{\Gamma} = \frac{\Gamma-2h\mathbb{1}_{2}}{1+h} \equiv \sigma_Z \tilde{\Sigma}_{A}\sigma_Z + \tilde{\Sigma}_{B} -\sigma_Z \tilde{\varepsilon}_{AB}-\tilde{\varepsilon}_{AB}^\intercal\sigma_Z.
\end{equation}
For the symmetric resource we define
\begin{eqnarray}\label{H2PS_sym_eff_submatrices}
\nonumber \tilde{\Sigma}_{A} &=& \frac{1}{1+h}\left(\Sigma_{A} - h\mathbb{1}_{2}\right), \\
\nonumber \tilde{\Sigma}_{B} &=& \frac{1}{1+h}\left(\Sigma_{B} - h\mathbb{1}_{2}\right), \\
\tilde{\varepsilon}_{AB} &=& \frac{1}{1+h}\varepsilon_{AB},
\end{eqnarray}
whereas for the asymmetric one we require $\tilde{\Sigma}_{A}=\tilde{\Sigma}_{B}$, such that
\begin{eqnarray}\label{H2PS_asym_eff_submatrices}
\nonumber \tilde{\Sigma}_{A} &=& \frac{1}{1+h}\left(\frac{\Sigma_{A}+\Sigma_{B}}{2} - h\mathbb{1}_{2}\right), \\
\nonumber \tilde{\Sigma}_{B} &=& \frac{1}{1+h}\left(\frac{\Sigma_{A}+\Sigma_{B}}{2} - h\mathbb{1}_{2}\right), \\
\tilde{\varepsilon}_{AB} &=& \frac{1}{1+h}\varepsilon_{AB}.
\end{eqnarray}
These ``re-Gaussified'' covariance matrices need to satisfy positivity and the uncertainty principle, meaning that $|\sqrt{\det\tilde{\Sigma}}-1|\geq |\tilde{\alpha}-\tilde{\beta}|$, assuming that we can write $\tilde{\Sigma}_{A}=\tilde{\alpha}\mathbb{1}_{2}$, $\tilde{\Sigma}_{B}=\tilde{\beta}\mathbb{1}_{2}$, and $\tilde{\varepsilon}_{AB} = \tilde{\gamma}\sigma_{Z}$. Furthermore, if $\Sigma_{A}=\alpha\mathbb{1}_{2}$, $\Sigma_{B}=\beta\mathbb{1}_{2}$, and $\varepsilon_{AB} = \gamma\sigma_{Z}$, this condition can be expressed as
\begin{equation}
\left|\sqrt{\det\Sigma}-h(2+\alpha+\beta)-1 \right| \geq (1+h)|\alpha-\beta|
\end{equation}
for the symmetric state, and as
\begin{equation}
\left|\sqrt{\det\Sigma}+\frac{1}{4}(\alpha-\beta)^{2} - h(\alpha+\beta) + h^{2} - 1 \right| \geq 0
\end{equation}
for the asymmetric one. A graphical proof that these conditions are met can be found in Appendix~\ref{app_D}.

\section{Probabilistic photon subtraction}\label{app_C}
In this section, we present the definitions we have used to shorten the notation of the modified submatrices of the covariance matrix of a Gaussian state that has undergone a symmetric two-photon-subtraction protocol (with beamsplitters and photodetectors). Assuming that $\Sigma_{A}$, $\Sigma_{B}$, and $\varepsilon_{AB}$ are symmetric, the general expression for the submatrices of the covariance matrix of the resulting state is
\begin{eqnarray}
\nonumber \tilde{\Sigma}_{A} &=& \tau \Sigma_{A} + (1-\tau)\mathbb{1}_{2} - 2\left( J_{1}X_{A}^{-1}J_{1}^{\intercal} + K_{1}Y^{-1}K_{1}^{\intercal}\right), \\
\nonumber \tilde{\Sigma}_{B} &=& \tau \Sigma_{B} + (1-\tau)\mathbb{1}_{2} - 2\left( J_{2}X_{A}^{-1}J_{2}^{\intercal} + K_{2}Y^{-1}K_{2}^{\intercal}\right), \\
\tilde{\varepsilon}_{AB} &=& \tau \varepsilon_{AB} - 2\left( J_{1}X_{A}^{-1}J_{2}^{\intercal} + K_{1}Y^{-1}K_{2}^{\intercal} \right),
\end{eqnarray}
whose success probability is given by
\begin{equation}
P = \frac{m_{1}m_{2}+m_{3}}{\sqrt{\det X_{A}\det Y}}. 
\end{equation}
Here, we have used
\begin{eqnarray}
\nonumber X_{A} &=& \frac{1}{2}\Omega \left[ (1-\tau)\Sigma_{A} + (1+\tau)\mathbb{1}_{2} \right]\Omega^{\intercal}, \\
\nonumber X_{B} &=& \frac{1}{2}\Omega \left[ (1-\tau)\Sigma_{B} + (1+\tau)\mathbb{1}_{2} \right]\Omega^{\intercal}, \\
\nonumber H &=& -\frac{1}{2}(1-\tau)\Omega \varepsilon_{AB}\Omega^{\intercal}, \\
\nonumber Y &=& X_{B} - HX_{A}^{-1}H, \\
\nonumber W_{X,M} &=& X^{-1}\tr(X^{-1}M) - \frac{\Omega M \Omega^{\intercal}}{\det X}, \\
\nonumber m_{1} &=& 1-\frac{1}{2}\tr Y^{-1}, \\
\nonumber m_{2} &=& 1-\frac{1}{2}\tr X_{A}^{-1} -\frac{1}{2}\tr\left(Y^{-1}HW_{X_{A},\mathbb{1}_{2}}H\right), \\
\nonumber m_{3} &=& \frac{1}{2}\tr\left(W_{Y,\mathbb{1}_{2}}HW_{X_{A},\mathbb{1}_{2}}H\right), \\
\nonumber K_{1} &=& \frac{1}{2}\sqrt{\tau(1-\tau)}\left[ \varepsilon_{AB}\Omega^{\intercal} + (\Sigma_{A}-\mathbb{1}_{2})\Omega^{\intercal}X_{A}^{-1}H \right], \\
\nonumber K_{2} &=& \frac{1}{2}\sqrt{\tau(1-\tau)}\left[ (\Sigma_{B}-\mathbb{1}_{2})\Omega^{\intercal} + \varepsilon_{AB}\Omega^{\intercal}X_{A}^{-1}H \right], \\
\nonumber J_{1} &=& \frac{1}{2}\sqrt{\tau(1-\tau)}(\Sigma_{A}-\mathbb{1}_{2})\Omega^{\intercal}, \\
J_{2} &=& \frac{1}{2}\sqrt{\tau(1-\tau)}\varepsilon_{AB}\Omega^{\intercal}.
\end{eqnarray}
The characteristic function of the resulting non-Gaussian state is
\begin{widetext}
\begin{eqnarray}
&& \chi^{(1,1)}(\alpha,\beta) = \frac{e^{-\frac{1}{4}\left[\vec{\alpha}\Omega \tilde{\Sigma}_{A}\Omega^{\intercal}\vec{\alpha}^{\intercal} + \vec{\beta}\Omega \tilde{\Sigma}_{B}\Omega^{\intercal}\vec{\beta}^{\intercal} + 2\vec{\alpha}\Omega \tilde{\varepsilon}_{AB}\Omega^{\intercal}\vec{\beta}^{\intercal} \right]}}{m_{1}m_{2}+m_{3}} \times \\
\nonumber && \left[ \left( m_{1} + \vec{\alpha}P_{1}\vec{\alpha}^{\intercal} + \vec{\beta}P_{2}\vec{\beta}^{\intercal} + \vec{\alpha}P_{12}\vec{\beta}^{\intercal} \right) \left(m_{2} + \vec{\alpha}Q_{1}\vec{\alpha}^{\intercal} + \vec{\beta}Q_{2}\vec{\beta}^{\intercal} + \vec{\alpha}Q_{12}\vec{\beta}^{\intercal} \right) + m_{3} + \vec{\alpha}R_{1}\vec{\alpha}^{\intercal} + \vec{\beta}R_{2}\vec{\beta}^{\intercal} + \vec{\alpha}R_{12}\vec{\beta}^{\intercal} \right].
\end{eqnarray}
\end{widetext}
Furthermore, when computing the average fidelity (Eq.~\eqref{eq:fidelity_2PS}), we obtain the non-Gaussian corrections defined by
\begin{widetext}
\begin{eqnarray}
\nonumber g &=& \frac{1}{m_{1}m_{2}+m_{3}}\bigg[ m_{1}\tr\left[\Omega\left(\mathbb{1}_{2}+\frac{1}{2}\tilde{\Gamma}\right)^{-1}\Omega^{\intercal}\left(ZQ_{1}Z + Q_{2} + ZQ_{12}\right)\right] + m_{2}\tr\left[\Omega\left(\mathbb{1}_{2}+\frac{1}{2}\tilde{\Gamma}\right)^{-1}\Omega^{\intercal}\left(ZP_{1}Z + P_{2} + ZP_{12}\right)\right] \\
\nonumber && +\tr\left[\Omega\left(\mathbb{1}_{2}+\frac{1}{2}\tilde{\Gamma}\right)^{-1}\Omega^{\intercal}\left(ZP_{1}Z + P_{2} + ZP_{12}\right)\right]\tr\left[\Omega\left(\mathbb{1}_{2}+\frac{1}{2}\tilde{\Gamma}\right)^{-1}\Omega^{\intercal}\left(ZQ_{1}Z + Q_{2} + ZQ_{12}\right)\right] \\
&& + \tr\left[\Omega\left(\mathbb{1}_{2}+\frac{1}{2}\tilde{\Gamma}\right)^{-1}\Omega^{\intercal}\left(ZR_{1}Z + R_{2} + ZR_{12}\right)\right] + 2\tr\left[ W_{\Omega\left(\mathbb{1}_{2}+\frac{1}{2}\tilde{\Gamma}\right)\Omega^{\intercal},ZP_{1}Z + P_{2} + ZP_{12}}\left(ZQ_{1}Z + Q_{2} + ZQ_{12}\right)\right] \bigg]
\end{eqnarray}
which can be computed using
\begin{eqnarray}
\nonumber P_{1} &=& -\frac{1}{2}\Omega K_{1}W_{Y,\mathbb{1}_{2}}K_{1}^{\intercal}\Omega^{\intercal}, \\
\nonumber P_{2} &=& -\frac{1}{2}\Omega K_{2}W_{Y,\mathbb{1}_{2}}K_{2}^{\intercal}\Omega^{\intercal}, \\
\nonumber P_{12} &=& -\Omega K_{1}W_{Y,\mathbb{1}_{2}}K_{2}^{\intercal}\Omega^{\intercal}, \\
\nonumber Q_{1} &=& -\frac{1}{2}\Omega\left( J_{1}W_{X_{A},\mathbb{1}_{2}}J_{1}^{\intercal} +2J_{1}W_{X_{A},\mathbb{1}_{2}}HY^{-1}K_{1}^{\intercal} + K_{1}W_{Y,HW_{X_{A},\mathbb{1}_{2}}H}K_{1}^{\intercal}\right)\Omega^{\intercal}, \\
\nonumber Q_{2} &=& -\frac{1}{2}\Omega \left(J_{2}W_{X_{A},\mathbb{1}_{2}}J_{2}^{\intercal} +2J_{2}W_{X_{A},\mathbb{1}_{2}}HY^{-1}K_{2}^{\intercal} + K_{2}W_{Y,HW_{X_{A},\mathbb{1}_{2}}H}K_{2}^{\intercal}\right)\Omega^{\intercal}, \\
\nonumber Q_{12} &=& -\Omega \left(J_{1}W_{X_{A},\mathbb{1}_{2}}J_{2}^{\intercal} + J_{1}W_{X_{A},\mathbb{1}_{2}}HY^{-1}K_{2}^{\intercal} + K_{1}Y^{-1}HW_{X_{A},\mathbb{1}_{2}}J_{2}^{\intercal}  + K_{1}W_{Y,HW_{X_{A},\mathbb{1}_{2}}H}K_{2}^{\intercal}\right)\Omega^{\intercal}, \\
\nonumber R_{1} &=& \frac{1}{2}\Omega\bigg[ J_{1} W_{X_{A},\mathbb{1}_{2}}HW_{Y,\mathbb{1}_{2}}K_{1}^{\intercal} \\
\nonumber &+& K_{1}\left(W_{Y,\mathbb{1}_{2}}\tr\left(Y^{-1}HW_{X_{A},\mathbb{1}_{2}}H\right) + Y^{-1}\tr\left(W_{Y,\mathbb{1}_{2}}HW_{X_{A},\mathbb{1}_{2}}H\right) - \frac{\Omega HW_{X_{A},\mathbb{1}_{2}}H\Omega^{\intercal}}{\det Y}\tr Y^{-1} \right) K_{1}^{\intercal} \bigg]\Omega^{\intercal}, \\
\nonumber R_{2} &=& \frac{1}{2}\Omega\bigg[ J_{2} W_{X_{A},\mathbb{1}_{2}}HW_{Y,\mathbb{1}_{2}}K_{2}^{\intercal} \\
\nonumber &+& K_{2}\left(W_{Y,\mathbb{1}_{2}}\tr\left(Y^{-1}HW_{X_{A},\mathbb{1}_{2}}H\right) + Y^{-1}\tr\left(W_{Y,\mathbb{1}_{2}}HW_{X_{A},\mathbb{1}_{2}}H\right) - \frac{\Omega HW_{X_{A},\mathbb{1}_{2}}H\Omega^{\intercal}}{\det Y}\tr Y^{-1} \right) K_{2}^{\intercal} \bigg]\Omega^{\intercal}, \\
\nonumber R_{12} &=& \frac{1}{2}\Omega\bigg[ J_{1} W_{X_{A},\mathbb{1}_{2}}HW_{Y,\mathbb{1}_{2}}K_{2}^{\intercal} + K_{1} W_{Y,\mathbb{1}_{2}}HW_{X_{A},\mathbb{1}_{2}} J_{2}^{\intercal} \\
\nonumber &+& 2 K_{1}\left(W_{Y,\mathbb{1}_{2}}\tr\left(Y^{-1}HW_{X_{A},\mathbb{1}_{2}}H\right) + Y^{-1}\tr\left(W_{Y,\mathbb{1}_{2}}HW_{X_{A},\mathbb{1}_{2}}H\right) - \frac{\Omega HW_{X_{A},\mathbb{1}_{2}}H\Omega^{\intercal}}{\det Y}\tr Y^{-1} \right) K_{2}^{\intercal} \bigg]\Omega^{\intercal}.
\end{eqnarray}

\end{widetext}

\section{Positivity and uncertainty principle for covariance matrices}\label{app_D}
\begin{figure}
\includegraphics[width=0.5 \textwidth]{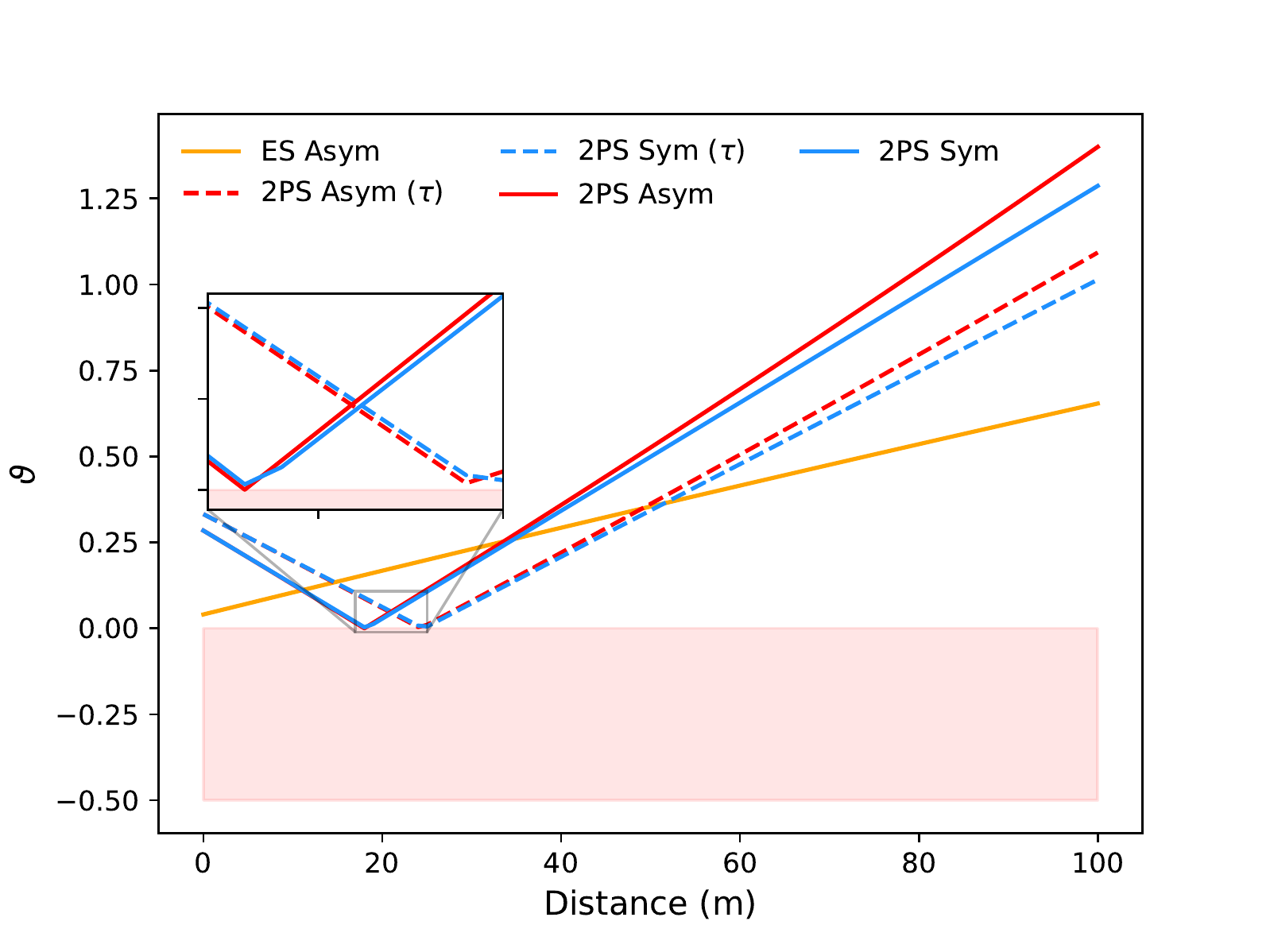}
\caption{Graphical representation of the curves constructed from Eq.~\eqref{check} against the travelled distance that, if positive, prove the submatrices used to compute this quantity characterize a covariance matrix, which satisfies a positivity condition, as well as the uncertainty principle. In orange, we represent the curve associated to the submatrices in Eq.~\eqref{ES_submatrices}, that result from entanglement swapping. The blue and red solid curves correspond to the heuristic two-photon-subtracted ``re-Gaussified'' symmetric and asymmetric states, respectively, described in Eqs.~\eqref{H2PS_sym_eff_submatrices} and~\eqref{H2PS_asym_eff_submatrices}. The blue and red dashed curves correspond to the probabilistic two-photon-subtracted ``re-Gaussified'' symmetric and asymmetric states, respectively, described in Eqs.~\eqref{2PS_sym_eff_submatrices} and~\eqref{2PS_asym_eff_submatrices}. As an inset, we investigate the region of short distances, in which we observe that the condition $\vartheta > 0$ is still met.}
\label{fig13}
\end{figure}
In this section, we discuss the conditions for the modified submatrices, product of the different entanglement distillation and entanglement swapping protocols discussed in the manuscript, to represent an actual covariance matrix of a different Gaussian state. The first one is positivity of the covariance matrix
\begin{equation}
\Sigma = \begin{pmatrix} \Sigma_{A} & \varepsilon_{AB} \\ \varepsilon_{AB}^{\intercal} & \Sigma_{B} \end{pmatrix} > 0,
\end{equation}
and the second one is preservation of the uncertainty principle,
\begin{equation}
\begin{pmatrix} \Sigma_{A} & \varepsilon_{AB} \\ \varepsilon_{AB}^{\intercal} & \Sigma_{B} \end{pmatrix} + i\begin{pmatrix} \Omega & 0 \\ 0 & \Omega \end{pmatrix} \geq 0.
\end{equation}
If we consider $\Sigma_{A}=\alpha\mathbb{1}_{2}$, $\Sigma_{B}=\beta\mathbb{1}_{2}$, and $\varepsilon_{AB}=\gamma\sigma_{Z}$, the positivity condition reduces to
\begin{eqnarray}
\alpha &>& 0, \\
\nonumber \det\Sigma &>& 0,
\end{eqnarray}
whereas the uncertainty principle can be written as
\begin{eqnarray}
\alpha &\geq& 1, \\
\nonumber \det\Sigma &\geq& \alpha^{2} + \beta^{2} - 2\gamma^{2} - 1.
\end{eqnarray}
Notice that the latter imposes a more restrictive condition. Given that any covariance matrix requires $\alpha\geq 1$ and $\beta\geq 1$, we can summarize all conditions into
\begin{equation}\label{check}
\vartheta \equiv \left|\sqrt{\det\Sigma} - 1\right| - \left|\alpha-\beta\right| \geq 0.
\end{equation}
In Fig.~\ref{fig13}, we investigate whether this condition is satisfied for different modified covariance matrices by representing $\vartheta$ against the travelled distance: in Eq.~\eqref{ES_submatrices}, entanglement swapping (orange); in Eqs.~\eqref{H2PS_sym_eff_submatrices} and~\eqref{H2PS_asym_eff_submatrices}, re-Gaussified heuristic photon subtraction (symmetric in a blue line, asymmetric in a red line); in Eqs.~\eqref{2PS_sym_eff_submatrices} and~\eqref{2PS_asym_eff_submatrices}, re-Gaussified probabilistic photon subtraction (symmetric in a blue dashed line, asymmetric in a red dashed line). Notice that all five cases satisfy both positivity and uncertainty principle conditions, confirming that they are indeed covariance matrices. As an inset, we zoom in the short distance behavior, where the curves approach the region in which $\vartheta < 0$, in a pale red background. As we can see, even in that area $\vartheta > 0$ is satisfied.


\end{document}